%% file: main.tex
\documentclass[acmsmall]{acmart}

\usepackage{kotex}
\usepackage{comment}
\usepackage{caption}
\usepackage{subcaption}
\newbool{showComments}
\booltrue{showComments}
\usepackage{colortbl}
\usepackage{color}
\ifbool{showComments}{%
\newcommand{\rev}[1]{{\color{black}{#1}}}

}
\AtBeginDocument{%
  \providecommand\BibTeX{{%
    \normalfont B\kern-0.5em{\scshape i\kern-0.25em b}\kern-0.8em\TeX}}}

\setcopyright{acmcopyright}
\copyrightyear{2018}
\acmYear{2018}
\acmDOI{XXXXXXX.XXXXXXX}

\acmConference[Conference acronym 'XX]{Make sure to enter the correct
  conference title from your rights confirmation emai}{June 03--05,
  2018}{Woodstock, NY}
\acmPrice{15.00}
\acmISBN{978-1-4503-XXXX-X/18/06}

\begin{document}

\title{Designing for Engaging Communication Between Parents and Young Adult Children Through Shared Music Experiences} 


\author{Euihyeok Lee}
\email{euihyeok.lee@misl.koreatech.ac.kr}
\affiliation{%
  \institution{Korea University of Technology and Education}
  \country{Republic of Korea}
}

\author{Souneil Park}
\email{souneil.park@nielseniq.com}
\affiliation{%
  \institution{NielsenIQ}
  \country{Spain}
}

\author{Jin Yu}
\email{jin.yu@misl.koreatech.ac.kr}
\affiliation{%
  \institution{Korea University of Technology and Education}
  \country{Republic of Korea}
}

\author{Seungchul Lee}
\email{seungchul@nclab.kaist.ac.kr;}
\affiliation{%
  \institution{KAIST}
  \country{Republic of Korea}
}

\author{Seungwoo Kang}
\authornote{Corresponding author}
\email{swkang@koreatech.ac.kr}
\affiliation{%
  \institution{Korea University of Technology and Education}
  \country{Republic of Korea}
}


\renewcommand{\shortauthors}{Lee, et al.}


\input{sections/00.Abstract}

\ccsdesc[500]{Human-centered computing~Collaborative and social computing systems and tools}
\ccsdesc[300]{Human-centered computing~Empirical studies in collaborative and social computing}

\settopmatter{printacmref=false}
\renewcommand\footnotetextcopyrightpermission[1]{}
\pagestyle{plain}

\keywords{Music-mediated interaction, parent-child communication, music sharing, music recommendation, mobile application, text messaging}


\maketitle

\input{sections/01.Introduction.tex}

\input{sections/02.RelatedWork.tex}

\input{sections/03.StudyProcedure.tex}

\input{sections/04.PreStudy.tex}
\input{sections/05-1.DesignStudy.tex}

\input{sections/05-2.Implementation.tex}

\input{sections/06.Evaluation.tex}

\input{sections/07.Discussion.tex}

\input{sections/08.Limitations.tex}

\input{sections/09.Conclusion.tex}

\bibliographystyle{ACM-Reference-Format}
\bibliography{reference}

\end{document}

%% file: sections/00.Abstract.tex
\begin{abstract}

This paper aims to foster social interaction between parents and young adult children living apart via music. Our approach transforms their music-listening moment into an opportunity to listen to the other's favorite songs and enrich interaction in their daily lives. To this end, we explore the current practice and needs of parent-child communication and the experience and perception of music-mediated interaction. Based on the findings, we developed DJ-Fam, a mobile application that enables parents and children to listen to their favorite songs and use them as conversation starters to foster parent-child interaction. From our deployment study with seven families over four weeks in South Korea, we show the potential of DJ-Fam to influence parent-child interaction and their mutual understanding and relationship positively. Specifically, DJ-Fam considerably increases the frequency of communication and diversifies the communication channels and topics, all of which are satisfactory to the participants.

\end{abstract}

%% file: sections/01.Introduction.tex
\section{Introduction}

Interaction between parents and children is essential to maintain a good parent-child relationship. The relationship quality is vital for psychological well-being in both generations~\cite{merz2009wellbeing, garcia2019role, xiang2020links, koropeckyj2002beyond}. Better family relationships help promote a higher level of psychological well-being of emerging adult children~\cite{garcia2019role}. High family cohesion can enhance the children's well-being~\cite{xiang2020links}. The high-quality relationship is also linked with a low level of loneliness and depression in parents~\cite{koropeckyj2002beyond}. 

However, parent-child interaction and relationships are reshaped after children leave home due to going to college or working in different cities or countries~\cite{shakeri2023sensing}, often involving changes in closeness in the relationship~\cite{golish2000changes}. If regular family communication between parents and young adult children living apart is not maintained, the parent-child relationship may be negatively impacted~\cite{smith2012going}. While they can use diverse communication methods such as phone calls, texting, and instant messaging to stay connected over a distance~\cite{mccurdy2022college, smith2012going}, it cannot always be possible to communicate with each other smoothly due to their duties and demanding work. Also, as the time children live apart increases, shared time spent together naturally decreases, resulting in a lack of shared experiences and contexts. It may lead to reduced communication frequency and limited communication topics, such as primarily focusing on short informational or logistical conversations. 

In this paper, we aim to explore the potential of music to foster communication between parents and young adult children living apart, focusing on establishing inter-generational empathy and creating opportunities for enriched daily interaction. Music is ubiquitous in our daily life, enjoyed by people of all ages and in various settings~\cite{park2021social}. Parents and their children tend to share similar music tastes~\cite{ter2011intergenerational}, and music transmitted from the parent generation to the child generation affects autobiographical memories, music preferences, and emotional responses to the music~\cite{krumhansl2013cascading}. It is also known that shared music-related activities positively affect family cohesion and emotional well-being~\cite{boer2014music}, and parent-child relationship quality~\cite{wallace2018associations}. Such characteristics of music in the family context might offer an excellent potential for enriched inter-generational interaction. 

Based upon previous research findings, we also note the situational characteristics of the moment of listening to music to take an opportunity for emotional and relational communication rather than a superficial and routine one. We often listen to music to relax or spend time free from urgency. It may provide a right-time opportunity for parents and children to understand each other's musical preferences and empathize. Also, it might be possible to leverage their mutual curiosity about each other's tastes to lead to rich interaction in their daily lives. Considering such opportunities, this paper seeks to address the following research question: How can we develop a mobile application to transform daily music-listening moments into opportunities for establishing inter-generational empathy and fostering rich interaction between parents and young adult children living apart?

To this end, we conducted a multi-phase study. Through a preliminary study with 21 participants, we explored the current communication practice between parents and young adult children living apart, their needs for communication, and their experience and perception of music-mediated parent-child interaction. Based on the findings of the preliminary study, we derived key features of the application: (1) it utilizes the moment users listen to music to create opportunities for establishing a shared context with the other generation while minimizing the hurdles and burdens of daily interaction. (2) it provides the users with recommended songs from the other generation's playlists, similar to the song they listen to, to help them develop a sense of empathy towards each other and be motivated to communicate. (3) it allows the users to share the recommended song with the other generation if they listen to it, making the other generation aware that the users have listened to it, which sparks their curiosity and naturally leads to an engaging conversation. We conducted a design study based on the derived features to evolve our initial concept into a concrete prototype. It allowed us to develop a feasible application that effectively works for real-world deployment. As a result, we implemented DJ-Fam, a mobile application that allows parents and children to listen to their favorite songs and facilitates conversation between them, using the songs as a conversation starter to feel empathy for each other.

We deployed DJ-Fam for four weeks with 14 participants, seven dyads of a parent and a young adult child living apart in South Korea. Through the deployment study, we examined the participants' experiences of DJ-Fam, their perception of using music for parent-child interaction, and the influence of DJ-Fam on their communication. The study results demonstrated the potential of DJ-Fam to foster enriched interaction between parents and their children. Specifically, we observe that the lightweight, asynchronous features of DJ-Fam indeed increase the communication frequency. Furthermore, the recommendations effectively raise curiosity and create the opportunity for richer conversation that goes beyond mundane topics and expands mutual understanding. 

The major contributions of this paper are as follows. First, we propose a way to foster social interaction between parents and young adult children living apart using their favorite songs as an opportunity for enriched conversation in their daily lives. Second, we explore the current practice of their communication, their needs, and the experience and perception of music-mediated parent-child interaction through interviews with 21 participants. Third, we develop DJ-Fam, a mobile application that transforms one's music-listening moment into an opportune moment for listening to each other's favorite songs and facilitates parents and children to use the songs as an engaging conversation topic. Fourth, we demonstrate the potential of DJ-Fam for enriching parent-child interaction based on the mutual appreciation of their favorite songs through a four-week deployment with seven families involving 14 participants.

%% file: sections/02.RelatedWork.tex
\section{Related work}

\subsection{Challenges and Opportunities of Social Interaction between Separated Families}
Separated families have a desire to stay connected with their loved ones and share updates about their recent lives and overall state~\cite{mynatt2001digital,neustaedter2006interpersonal, romero2007connecting}. However, they face several challenges that make their social interaction difficult, and many studies have explored these issues. 
Key challenges identified in previous studies include difficulties in coordinating schedules among family members~\cite{modlitba2008globetoddler, tee2009exploring, judge2011familyportals, heshmat2020familystories}, generational gaps in technological proficiency~\cite{tee2009exploring, wallbaum2018supporting, ballagas2009family, yuan2016they}, the psychological burden associated with initiating or maintaining contact~\cite{kirk2006understanding,yarosh2009supporting,kang2021momentmeld}, and concerns regarding privacy intrusion~\cite{yin2024methodological,judge2011familyportals,yarosh2009developing}.

Various forms of separated families experience unique challenges or needs depending on their relationships and specific circumstances. 
Married couples living apart often struggle with adjusting work schedules for communication~\cite{bunker1992quality,priastuty2023long}. They also need to address the lack of intimate interactions that were present when together, as they share a close emotional and physical bond~\cite{bales2011couplevibe}. Grandparent-grandchild relationships often face technological gaps that hinder communication~\cite{ballagas2009family,yuan2016they,wallbaum2018supporting}, and differences in content preferences could limit engagement~\cite{olsson2008user}. Some studies focusing on siblings emphasized the need for designs that accommodate their long-lasting relationships, which can be hierarchical but also seek equal and comfortable interactions~\cite{jin2023socio}. Parent-child interactions have been extensively studied, spanning different life stages such as young parents with children, emerging adult children and their parents, and elderly parents with adult children. These relationships present challenges such as differing expectations, lack of shared interests, and emotional communication difficulties~\cite{shin2021designing}.

In this paper, we focus on the relationship between emerging adult children and their parents living apart and seek technological opportunities to enhance their interaction. Their relationship shares some similarities with other family relationships, but also has a few unique characteristics. First, this stage marks a significant transition as children who once lived with their parents move out, leading to the need for new forms of communication and emotional exchange~\cite{shakeri2023sensing,golish2000changes,flanagan1993residential}. In particular, emerging adults tend to focus on themselves, prioritizing academics, social life, and personal development~\cite{arnett2001adolescence}, often resulting in a preference for concise communication~\cite{yarosh2011mediated}. Excessive parental interference can have negative effects on children~\cite{lindell2017implications}, prompting parents to seek communication methods that allow them to maintain an appropriate level of distance as their children transition into adulthood~\cite{smith2012going}. At the same time, emerging adults are often not yet fully independent from their parents, making parental support essential~\cite{nice2023features}. They also experience instability due to various life transitions~\cite{arnett2020features}, leading to a continued desire for regular communication driven by mutual concern and care~\cite{arnett2000emerging,hofer2011iconnected}. Thus, their relationship cannot be solely defined by parental authority and protection, or by the child's independence. Instead, it requires communication that balances mutual understanding of roles, appropriate distance, and emotional exchange. However, previous research remains significantly lacking in exploring opportunities for this relationship and providing design solutions that practically address their distinct challenges and needs.

\subsection{Tools for Bonding between Remote Family Members}

Our work relates to previous studies aimed at creating a sense of connectedness to foster bonding between separated families in a broader socio-technical context.
In this broader context, we draw upon literature on virtual presence and co-activity systems that share similar goals. Unlike DJ-Fam, with which users share music listening experiences asynchronously, much of the literature primarily addresses temporally synchronized connections. We review the works in two categories: systems that establish a persistent presence and those that support specific activities. 

\textit{Media space}~\cite{bly1993media} was a pioneering effort and has inspired much research on virtual connections between physical spaces and co-presence. Originally, media spaces aimed to support remote collaboration by creating a synchronous, persistent audio and video connection, or always-on video, between two distant workplaces~\cite{mantei1991experiences}. 
Many researchers have extended the use of media spaces beyond workplaces to connect distant family members. These studies have often adopted the concept of always-on audio or video~\cite{hindus2001casablanca, judge2010family, judge2011familyportals, tang2013homeproxy}. Some variations are also explored; for example, HomeProxy~\cite{tang2013homeproxy} mixing synchronous and asynchronous videos, and the use of shared whiteboards as secondary interfaces for playful interactions, such as in CommuteBoard~\cite{hindus2001casablanca}, Family Window~\cite{judge2010family}, and Family Portals~\cite{judge2011familyportals}. 
Baishya \& Neustaedter extend the scope of media space into everyday life beyond a home using a wearable camera for long-distance couples~\cite{baishya2017inyoureyes}. These studies highlight the benefits of always-on media for connectedness between family members or loved ones, but also report privacy concerns, as users often feel uneasy about exposing private spaces to others~\cite{baishya2017inyoureyes, judge2010family}.

Co-activity systems have also been developed to study interactions between young children and their parents or grandparents, in activities such as book reading~\cite{follmer2012people, raffle2011hello, kang2017zaturi}, gaming~\cite{yarosh2009developing, yarosh2013almost}, and music listening~\cite{tibau2019familysong}.
Focusing on specific activities helps to minimize privacy concerns associated with providing persistent awareness. Additionally, these systems are effective in engaging young children, as they offer short-term activities suited to children’s limited attention spans~\cite{ballagas2012reading, modlitba2008globetoddler}. 
Though DJ-Fam is not temporally synchronous, it functions as a music-listening support system for families. 
Our work studies new communication opportunities unexplored in synchronized co-listening systems, which could be substantial given the challenge of finding suitable times for parents and young adult children living apart.

\subsection{Music for Fostering Social Interaction}

This study focuses on social interaction mediated by music, aiming to enhance the mutual understanding and relationship between parents and young adult children who live apart. Music is one of the most fundamental tools for expressing and communicating one's identity and emotions~\cite{schafer2020music, north1999music}. Many studies in psychology have explored facilitating social interaction as one of the core functions of music~\cite{lonsdale2011we, schafer2009functions, boer2012towards}. These studies emphasize that music plays a vital role in helping individuals form and maintain social relationships, such as ice-breaking with strangers~\cite{rentfrow2006message}, socializing with friends~\cite{lonsdale2011we}, and the formation of bonds~\cite{boer2011shared, brown2001music}. It has also been reported that sharing music in various forms (e.g., posting links to music on social media) can catalyze social interaction, encouraging dialogue and story sharing~\cite{leong2013revisiting, park2022cross}.

Beyond general relationships, music-related activities have also been found to increase understanding and improve relationships among family members. Wallace and Harwood reported that shared musical engagement between parents and children in adolescence, such as listening to music and talking about favorite music, positively impacts interpersonal coordination and cognitive empathy, subsequently affecting relational quality~\cite{wallace2018associations}. Similarly, Boer and Abubakar showed that musical family rituals (e.g., talking to one's family about music and listening to music with family) positively affect family cohesion, providing an easy and enjoyable way to maintain family bonds~\cite{boer2014music}.

Based on these social properties of music, various tools for music sharing and music-mediated interaction have been developed in HCI/CSCW literature. We review the literature by focusing on two factors. The first is the group of people the tools connect, ranging from the general public to family members, as this influences the tools' context and design. The second is the tool's role, which varies from basic functions such as discovery and connection of peers to more active roles, including music taste analysis and recommendation. 

With the advancement of mobile devices and connectivity, a branch of research aimed to transform `personal' music consumption (e.g., mp3 players) into `shared' experiences with nearby people and explore its impact on sociality. tunA ~\cite{bassoli2006tuna} and Push!Music ~\cite{hakansson2007facilitating} are early examples, enabling music sharing among mobile devices within a certain range. Kirk et al. developed PocketSong ~\cite{kirk2016understanding}, which shares a similar foundational idea of connecting people in a specific location. However, it uses physical space as an anchor, allowing music sharing between nearby users and those who previously visited or will visit the space. Jukola~\cite{ohara2004jukola} also links individuals within a shared public place, such as a bar, with the unique goal of facilitating collective playlist creation for a communal listening experience. As these tools connect spontaneous groups by location, often users with no prior relationship, their primary role lies in peer discovery or networking for sharing music. Some tools include social features, such as Jukola's voting and moderation mechanisms and tunA's messaging interface. However, music selection—the key feature of music sharing—still relies on users’ manual choices.

Social Playlist~\cite{liu2008social} and Colisten~\cite{stewart2018co} focus on social groups with established relationships, such as friends. Both systems aim to support a shared listening experience for group members at a distance, as if they were collectively tuning in to a shared radio channel. Accordingly, their primary role is to connect group members and facilitate a synchronized listening experience. While these systems offer moderation features for managing playlists and group memberships, their designs adopt user-driven playlist creation, remaining dependent on users’ input for music selection.

The closest branch of related works focuses on connecting remote family members through synchronized listening experiences. FamilySong~\cite{tibau2019familysong} was designed to facilitate communication among three generations of distant family members (i.e., grandparents, parents, and children). Chowdhury et al.~\cite{chowdhury2021listening, chowdhury2022music} examined the feasibility of using existing video conferencing and music streaming platforms to enable grandparents and grandchildren living far apart to listen to music together and engage in conversation. Again, the primary role of these tools is to support synchronized listening; for instance, FamilySong employs a specialized music playback device developed and deployed within family members' homes. Similar to our work, these studies emphasize the impact on family relationships, showing that co-listening sessions are experienced as a form of gathering, facilitating intimate and affective interactions, while transmitting cultural values and traditions from older to younger generations.

Prior research has addressed the potential of music-mediated tools for social connection, showing the positive impact of shared music-listening experiences on social interaction and bonding. However, these approaches largely rely on real-time participation for synchronized co-listening, often requiring simultaneous engagement or active coordination for interaction with friends and family. DJ-Fam shares the objective of strengthening social connection for family bonds and builds on the understanding that music sharing fosters intimate communication. Our work extends the literature by taking a different approach, targeting parent-emerging adult child relationships. In DJ-Fam, focusing on this specific social group not only sets the context but also directly informs the design and key features. Specifically, it focuses on the asynchronous sharing of music-listening experiences, which naturally blends into users' existing routines—namely, their everyday music-listening moments. We believe the approach well addresses difficulties in coordinating schedules, the burden of initiating and sustaining communication, and privacy concerns. Moreover, it better accommodates the needs of parent-emerging adult child relationships, such as the need for emotionally rich yet concise interaction and the careful balance between closeness and independence. Our deployment study further contributes to the literature by providing findings that illustrate how the tool effectively builds bridges between family members and fosters intimate communication for our target groups, which were not addressed before.

%% file: sections/03.StudyProcedure.tex
\section{Study Procedure}

\begin{figure}[h!]
    \centering
    \begin{minipage}{0.64\textwidth}
        \centering
        \includegraphics[width=\linewidth]{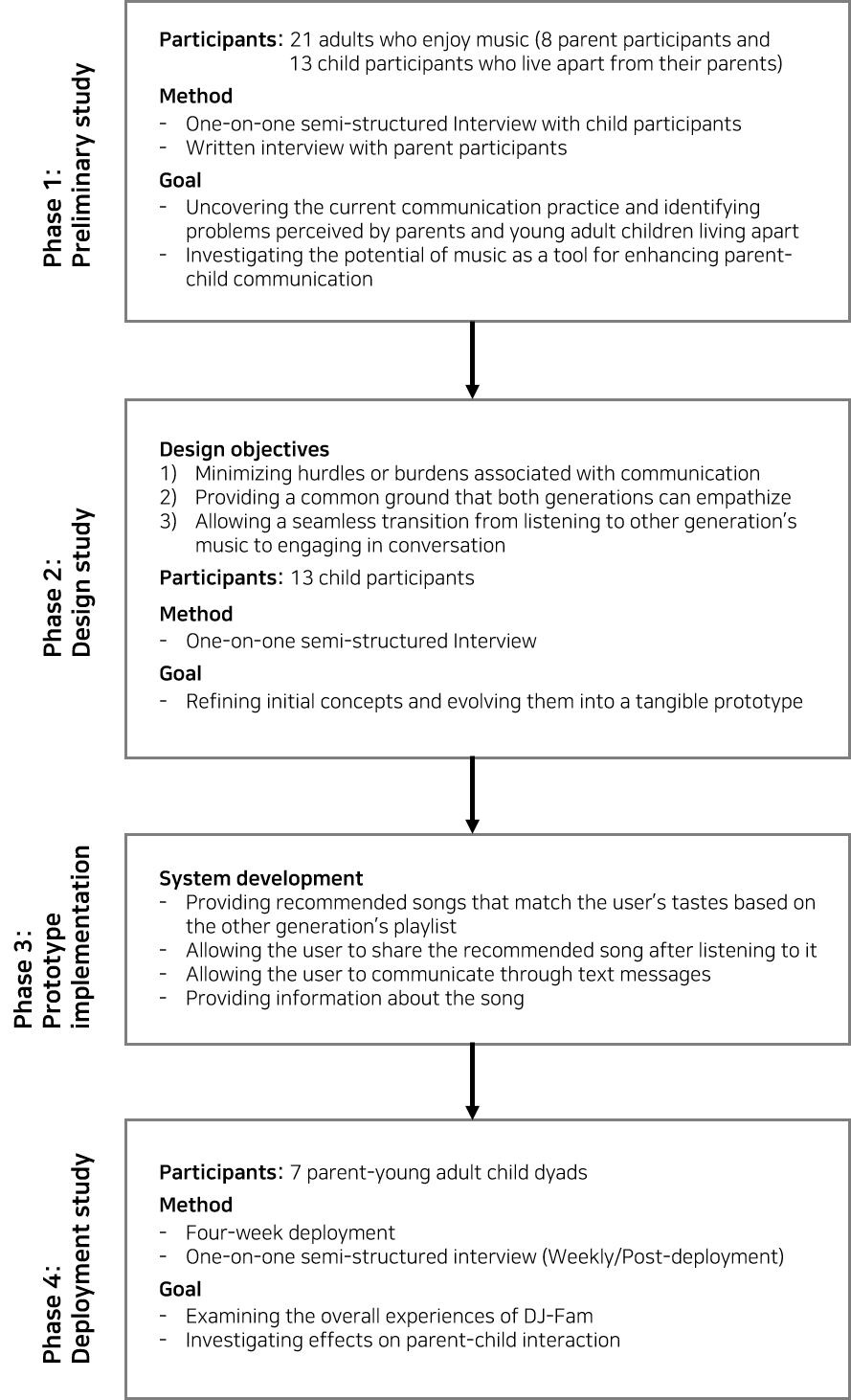}
        \caption{Study procedure~\label{fig:Study_procedure}}
        
    \end{minipage}\hfill
    \begin{minipage}{0.36\textwidth}
        \centering
        \begin{tabular}{|ccc|}
            \hline
            \rowcolor{lightgray}
            \multicolumn{3}{|c|}{\textbf{Parent generation}}                                                             \\ \hline
            \multicolumn{1}{|c|}{\textbf{ID}} & \multicolumn{1}{c|}{\textbf{Age}} & \multicolumn{1}{c|}{\textbf{Gender}} \\ \hline
            \multicolumn{1}{|l|}{P\_P1 \ \ $\dag$$\diamondsuit$}          & \multicolumn{1}{c|}{51}           & \multicolumn{1}{c|}{F}           \\ \hline   
            \multicolumn{1}{|l|}{P\_P2 \ \ $\dag$$\diamondsuit$}          & \multicolumn{1}{c|}{56}           & \multicolumn{1}{c|}{F}            \\ \hline
            \multicolumn{1}{|l|}{P\_P3 \ \ $\dag$$\diamondsuit$}          & \multicolumn{1}{c|}{51}           & \multicolumn{1}{c|}{M}            \\ \hline
            \multicolumn{1}{|l|}{P\_P4 \ \ $\dag$$\diamondsuit$}          & \multicolumn{1}{c|}{52}           & \multicolumn{1}{c|}{F}               \\ \hline
            \multicolumn{1}{|l|}{P\_P5 \ \ $\dag$$\diamondsuit$}          & \multicolumn{1}{c|}{53}           & \multicolumn{1}{c|}{F}               \\ \hline
            \multicolumn{1}{|l|}{P\_P6 \ \ \ \ $\diamondsuit$}          & \multicolumn{1}{c|}{51}           & \multicolumn{1}{c|}{F}               \\ \hline
            \multicolumn{1}{|l|}{P\_P7 \ \ \ \ $\diamondsuit$}          & \multicolumn{1}{c|}{55}           & \multicolumn{1}{c|}{F}               \\ \hline
            
            \multicolumn{1}{|l|}{P\_P8 \ \ $\dag$ }          & \multicolumn{1}{c|}{49}           & \multicolumn{1}{c|}{F}               \\ \hline
            \multicolumn{1}{|l|}{P\_P9 \ \ $\dag$}          & \multicolumn{1}{c|}{52}           & \multicolumn{1}{c|}{M}               \\ \hline
            \multicolumn{1}{|l|}{P\_P10 $\dag$}          & \multicolumn{1}{c|}{55}           & \multicolumn{1}{c|}{M}                   \\ \hline
            
            \multicolumn{3}{l}{ }            \\  
            
            \hline
            \rowcolor{lightgray}
             \multicolumn{3}{|c|}{\textbf{Child generation}}                                                             \\ \hline
             \multicolumn{1}{|c|}{\textbf{ID}} & \multicolumn{1}{c|}{\textbf{Age}} & \textbf{Gender}                         \\ \hline
             \multicolumn{1}{|l|}{P\_C1 \ \ $\dag$$\ddag$$\diamondsuit$}          & \multicolumn{1}{c|}{25}           & M               \\ \hline
             \multicolumn{1}{|l|}{P\_C2 \ \ $\dag$$\ddag$$\diamondsuit$}          & \multicolumn{1}{c|}{25}           & M               \\ \hline
             \multicolumn{1}{|l|}{P\_C3 \ \ $\dag$$\ddag$$\diamondsuit$}          & \multicolumn{1}{c|}{23}           & F               \\ \hline
             \multicolumn{1}{|l|}{P\_C4 \ \ $\dag$$\ddag$$\diamondsuit$}          & \multicolumn{1}{c|}{23}           & F               \\ \hline
             \multicolumn{1}{|l|}{P\_C5 \ \ $\dag$$\ddag$$\diamondsuit$}         & \multicolumn{1}{c|}{24}           & F               \\ \hline
             \multicolumn{1}{|l|}{P\_C6 \ \ \ \ \ \ $\diamondsuit$}         & \multicolumn{1}{c|}{30}           & M               \\ \hline
             \multicolumn{1}{|l|}{P\_C7 \ \ \ \ \ \ $\diamondsuit$}         & \multicolumn{1}{c|}{27}           & F               \\ \hline
             \multicolumn{1}{|l|}{P\_C8 \ \ $\dag$$\ddag$}         & \multicolumn{1}{c|}{24}           & F               \\ \hline
             \multicolumn{1}{|l|}{P\_C9 \ \ $\dag$$\ddag$}         & \multicolumn{1}{c|}{24}           & M               \\ \hline
             \multicolumn{1}{|l|}{P\_C10 $\dag$$\ddag$}         & \multicolumn{1}{c|}{30}           & M               \\ \hline
             \multicolumn{1}{|l|}{P\_C11 $\dag$$\ddag$}         & \multicolumn{1}{c|}{24}           & F               \\ \hline
             \multicolumn{1}{|l|}{P\_C12 $\dag$$\ddag$}         & \multicolumn{1}{c|}{23}           & F               \\ \hline
              \multicolumn{1}{|l|}{P\_C13 $\dag$$\ddag$}         & \multicolumn{1}{c|}{25}           & M               \\ \hline
             \multicolumn{1}{|l|}{P\_C14 $\dag$$\ddag$}         & \multicolumn{1}{c|}{23}           & F               \\ \hline
             \multicolumn{1}{|l|}{P\_C15 $\dag$$\ddag$}         & \multicolumn{1}{c|}{23}           & M               \\ \hline
            
             \multicolumn{3}{l}{ }            \\              
            \multicolumn{3}{l}{$\dag$: Preliminary study }            \\          
            \multicolumn{3}{l}{$\ddag$: Design study}                     \\
            \multicolumn{3}{l}{$\diamondsuit$: Deployment study}    \\
            
            \end{tabular}%
        \vspace{0.1in}
        \captionof{table}{Participant demographics}
        \label{tab:pre-study-demo}
    \end{minipage}
\end{figure}

To develop DJ-Fam, we designed the study procedure with a goal-oriented approach. Specifically, the study focused on exploring the potential of music in expanding and fostering opportunities for parent-child interaction in a way that seamlessly integrates into everyday life. We also sought to realize such opportunities in real-world settings by considering the users' needs and preferences. The study was performed along three phases of user studies, followed by a development phase, as shown in Figure~\ref{fig:Study_procedure}. All studies were carried out with IRB approval, and before each study, we explained the procedures and obtained informed consent from participants. We performed all user studies in South Korea.

\textbf{Preliminary study:} We explored the current practices between parents and young adult children living apart and examined the potential of using music to encourage their interaction. For this study, we recruited 21 participants (eight parent-child dyads and five additional child participants), as outlined in Table~\ref{tab:pre-study-demo}. Recruitment was conducted via word-of-mouth and snowball sampling methods. The child participants had begun living apart from their parents after starting college. While all child participants reported using music streaming services to listen to music, most parents preferred using YouTube. One parent also used the radio as an additional channel. 

We conducted one-on-one, semi-structured interviews with the child participants via video conferencing, each lasting 30-60 minutes. However, due to scheduling challenges with the parent participants, we opted for written interviews. The questionnaire comprised key questions along with several follow-up questions designed to delve deeper into the reasons or specific examples. We asked the parents to provide detailed written responses.

The key questions included: How often do you contact your child/parent? Are you satisfied or dissatisfied with the current frequency of contact, and why? How do you typically communicate with your child/parent? What topics do you usually discuss? Can you share any experiences of enjoying music with your family? What were they like if you've ever had conversations about music-related topics with your child/parent? How familiar are you with your parent/child's musical preferences and the music of their generation? After the interviews, participants were compensated with a gift voucher worth approximately USD 8. Section~\ref{sec:preliminary} presents the results of the preliminary study.

\textbf{Design study:} We devised an initial concept of DJ-Fam and derived its key features from preliminary study findings. We conducted a design study to transform these features into a concrete prototype. We invited 13 child participants from the preliminary study to participate in this phase (P\_C1 to P\_C5, P\_C8 to P\_C15 in Table~\ref{tab:pre-study-demo}). Considering their high familiarity with various mobile services such as music streaming, instant messaging, and social media apps, we focused on the younger generation. Since these participants were likely to have similar experiences with the app being studied, they were better positioned to understand its usage and interaction patterns, allowing them to provide more meaningful feedback on the app's features and user interface. To facilitate their understanding, we prepared two mock-up app screens before the study to help participants envision the features and UI.

We conducted one-on-one semi-structured interviews via video conferencing, each lasting between 30 minutes to an hour. Our objective was to explore the following questions: What would be a suitable format for users to communicate using music as a conversation topic? How should recommended music be presented, and how can the status of listening to the recommended songs be shared across generations? What type of music-related information would help facilitate communication? During the interviews, we showed and explained the mock-up app screens to participants and gathered their thoughts and opinions. The design study results are presented in Section~\ref{subsec:design_study}.

\textbf{Application prototype implementation:} Building on the features derived from the design study and user feedback, we developed a prototype of DJ-Fam on the Android platform. The app integrates instant messaging and music streaming functionalities. While DJ-Fam could support any type of music, we specifically targeted a specific type of music, i.e., song (vocal music with lyrics), to simplify the prototype development and the deployment study. 
For ease of use, we designed the interface similar to popular mobile apps that users already know. DJ-Fam recommends songs from the other generation's playlists, matching them with songs users are currently listening to. Users can share the recommended song via instant messaging, potentially sparking conversations about the music. The details of the DJ-Fam prototype implementation are provided in Section ~\ref{sec:implementation}

\textbf{Deployment study:} We conducted a four-week deployment study with seven families (i.e., 14 participants, consisting of seven parent-child dyads) to explore user experiences with DJ-Fam. Five of the dyads from the preliminary study were invited to participate (P\_P1 to P\_P5 and P\_C1 to P\_C5 in Table~\ref{tab:pre-study-demo}). We recruited two new dyads specifically for the deployment study through word-of-mouth (P\_P6 and P\_C6, P\_P7 and P\_C7 in Table~\ref{tab:pre-study-demo}).

Before starting the study, we asked participants questions to compare and understand the differences between their responses before and after the study. For the two newly recruited dyads, we first posed the same questions as in the preliminary study. We then inquired with all the participants about how often they communicated using other methods, such as phone calls or messaging apps (referred to as 'Baseline'), during the week before the deployment. After the deployment, we again asked about the frequency of communication using both DJ-Fam and Baseline methods.

To prepare the DJ-Fam app, we asked the participants to select their 100 favorite songs from 1980 to 2021 to create their playlists for the study. This time frame was chosen considering the participants' ages, particularly the parents'. Utilizing the list of 100 songs provided by each participant, we prepared and stored song-related data in the database used by the app, and DJ-Fam offered these songs during the deployment. Since we developed the prototype app for Android, we provided an app installation file directly to participants using Android smartphones rather than distributing it through the app market. Participants downloaded and installed the file on their devices, with child participants assisting their parents when they encountered difficulties downloading and installing the file from an unknown source. For iPhone users, we provided Android smartphones with the app pre-installed. Additionally, we supplied all participants with a manual explaining the app's usage.

The participants freely used DJ-Fam for four weeks. We conducted weekly intermediate interviews during the study, followed by post-deployment interviews after four weeks of usage. All interviews were one-on-one semi-structured sessions, conducted face-to-face for some participants and via phone for others who faced geographical constraints. Each interview lasted approximately 30 minutes to one hour. Before the interviews, we emphasized that the confidentiality of personal information, such as text messages, would be strictly maintained. We investigated their usage patterns, satisfaction with DJ-Fam, and perceived changes in communication and relationships between parents and their children. Each family was compensated with a gift voucher worth about USD 80 for participation.

During the weekly interviews, we asked participants about app usage and explored their feelings toward DJ-Fam. The questions included: What issues or difficulties did you encounter while using DJ-Fam, if any? When and how many times did you use the app? What are your thoughts on the recommended songs? What conversations did you have regarding those songs? Were there instances of communication where DJ-Fam was not utilized? Did you notice any changes compared to the previous week?

In the post-deployment interview, we investigated participants' overall experiences, satisfaction, and perceived changes after using DJ-Fam for four weeks. We asked questions about these aspects in addition to those posed during the weekly interviews. The questions included: When and where did you primarily use DJ-Fam? How satisfied or dissatisfied are you with the app? How do you feel about your communication and relationship with your parent/child while using DJ-Fam? What perceived changes, if any, did you notice after using the app? We present our findings from the deployment study in Section~\ref{sec:deployment_study}.

\subsection{Data Collection and Analysis}
Our study involves interview sessions in Phases 1, 2, and 4 (shown in Table~\ref{tab:pre-study-demo}) of the study procedure. Generally, our interview questions were designed to be open-ended and non-judgmental~\cite{kuniavsky2003observing}. We also took care to minimize any potential discomfort for the participants. Before starting each interview, we consistently assured participants that their anonymity would be strictly preserved, emphasized that there were no right or wrong answers, and encouraged them to speak freely. We listened attentively without interruption until they finished talking to ensure they felt comfortable responding. When participants gave general answers, we followed up with specific questions to elicit more detailed examples and authentic thoughts. Additional care was taken during Phases 1 and 4 interviews, as these could touch on more sensitive or private topics. At the start of these interviews, we reiterated our commitment to confidentiality, and we refrained from pressing further when participants had no response or seemed hesitant to share more.

To analyze the interviews, we audio-recorded and transcribed them. The written responses from the parent participants who completed interviews in writing were also included in the transcribed data. Two researchers participated in Phases 1 and 2 for data analysis, and three were involved in Phase 4. We employed a thematic analysis approach to investigate underlying themes~\cite{braun2006using}. The researchers engaged in each phase, reading and listening to the data multiple times to immerse themselves fully in the content. Initial impressions and ideas were noted to guide the subsequent analysis. Then, the researchers coded the interviews line by line independently, capturing interesting features across the dataset. After the initial coding, the codes were aggregated to identify potential themes. The researchers critically reviewed these themes to ensure coherence and relevance. Through iterative discussions, they reached a consensus on the specifics of each theme, assigning clear definitions and names to ensure the themes effectively encapsulated the underlying content. Note that the participants' interviews quoted in the paper are translated into English for presentation.

%% file: sections/04.PreStudy.tex
\section{Preliminary study}
\label{sec:preliminary}

\rev{Our goal of the preliminary study was to uncover the current communication practices and identify problems perceived by parents and their young adult children living apart. Moreover, we sought to investigate the potential of music as a tool for enhancing parent-child communication. In the following, we present findings from the interviews with 21 participants, including 8 parents and 13 children.}

\subsection{Current Communication Practices}
Participants reported an average of 4.3 instances of communication per week with their children or parents (parents: 4.75, children: 4), with all participants stating they communicated at least once a week. Six participants reported the highest frequency of communication, at seven times a week, typically once a day. Those who perceived their communication as infrequent offered different primary reasons. Parents believed their children might feel burdened or be busy with work or studies. On the other hand, children cited reasons such as lack of time, personal matters, only reaching out when necessary, or feeling that there was nothing specific to discuss.

They reported that they primarily communicated via phone calls rather than text messages. These conversations typically revolved around mundane and routine topics, such as whether they had eaten, how they were doing, plans to visit the parents' home, or other practical matters. Due to spending less time together, participants felt they had limited shared experiences to discuss, often resulting in brief conversations. Additionally, many parents noted a tendency for communication to feel one-sided, focusing on their children. Parents would ask questions in these cases, and the children would respond. However, most child participants did not perceive the communication similarly. They believed that communication with their parents is balanced.



\subsection{Needs for Improved Communication}
Most participants expressed a desire for more communication with their children/parents, though the two groups differed in the degree of need. Seven out of eight parents (88\%) expressed a desire for more frequent communication, with 43\% indicating that even daily contact felt insufficient. These parents wanted to share more of their experiences and better understand their children's thoughts, mainly since they lived apart. They also noted that increased communication would help them gain deeper insights into their children's lives.

On the other hand, eight out of thirteen child participants (62\%) expressed a need for more communication. They cited similar reasons, noting that more communication was important because they didn't know their parents' status well enough and lived far from them. Three participants mentioned that while they had sufficient communication with one parent, they lacked meaningful conversation with the other, often due to a lack of shared topics or empathy. Meanwhile, five child participants reported being satisfied with their current communication frequency—averaging three times per week—and felt no need for additional contact.

 \subsection{Experience of Listening to Music Together with Family Members}

Most participants reported having shared experiences of listening to music with their families, often enjoying these moments while traveling by car. However, now that they live apart, such opportunities have become less frequent.

When listening to music with their families, the younger generation usually chose songs. Possible misunderstandings were observed between the children and their parents. The children mentioned they considered their parents' musical tastes when selecting tracks, assuming they knew their parents' preferences and the songs from their generation well. In contrast, most parents reported that, although the songs played often do not match their tastes, they usually just listened to them anyway. 

Interesting differences also emerged regarding the awareness of each other's music preferences. As noted, children assumed they understood their parents' musical tastes but generally believed their parents would not know their preferences. All parents acknowledged their limited knowledge of songs from their children's generation, although some (six out of eight) mentioned being aware of their children's specific preferences. Four out of eight parents stated they tried to engage with their children's music because it was what their children enjoyed and to foster understanding, communication, and empathy.

\subsection{Potential of Music to Foster Parent-Child Dialogue}

Nineteen out of 21 (90\%) expressed a favorable attitude toward using music that appeals to both generations as a conversation starter to enhance communication with their parents or children. They cited various reasons, including the potential to understand each other's preferences better, foster empathy, and ease of conversation around the topic.
In particular, five parents highlighted the value of sharing memories tied to the music they enjoyed in the past, as it holds a special place in their hearts. Conversely, one child participant, who was not as positive, mentioned that he did not like the genre of music his parents preferred, making it challenging to envision discussing those songs. However, he noted that he might be open to conversation if there were common favorite genres.

14 out of 21 (67\%) reported having discussed music-related topics with their families, such as opinions on songs, artists, and the popularity of certain tracks. Conversations also touched on the times when certain songs were popular and their presence in movies, TV shows, and dramas. Some parents recalled their youth while sharing old songs with their children. 

%% file: sections/05-1.DesignStudy.tex
\section{Design Study}
\label{sec:design}

\subsection{Initial DJ-Fam Concept}







We derive several key considerations for DJ-Fam design based on findings from our formative study. First, lowering the barrier to communication is essential—not only due to the heightened need and expectation for interaction between family members but also because of the complexity introduced by living apart. Second, while music has the potential to facilitate communication, the most effective songs will be those that resonate emotionally with both generations. Lastly, there should be a natural transition from music listening to conversation, with a mechanism that encourages this shift.


The initial concept of DJ-Fam was designed to include the following features accordingly: (1) a single app where users can listen to music and chat asynchronously, allowing flexible adaptation to their individual situations. (2) recommendations based on the other generation's playlists, fostering empathy and conversation; (3) leveraging the moment of music listening to create shared context; and (4) the ability to share songs with the other generation, which could spark curiosity and initiate conversations.


\subsection{Design Study Results~\label{subsec:design_study}}

The design study closely examined user needs and preferences to refine the initial design into a functional prototype. In the following, we present results from interviews with 13 participants.

\subsubsection{What would be an appropriate format for users to communicate through music?}

\begin{figure}
     \centering
     \begin{subfigure}[b]{0.4\textwidth}
         \centering
         \includegraphics[width=1\textwidth]{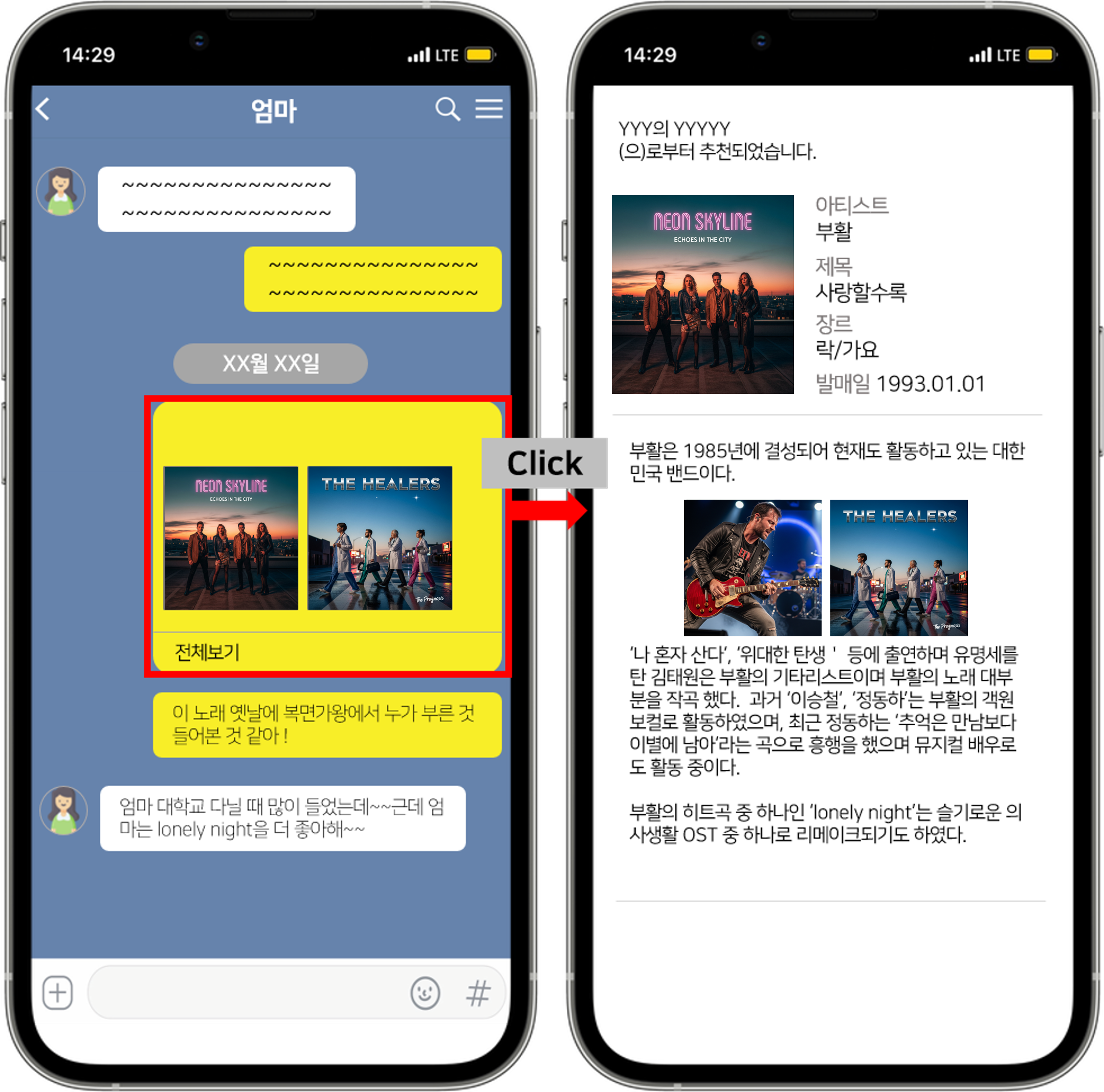}
         \caption{Messenger format}
         \label{fig:storyboard_messenger}
     \end{subfigure}
     \hspace{0.3in}
     \begin{subfigure}[b]{0.4\textwidth}
         \centering
         \includegraphics[width=1\textwidth]{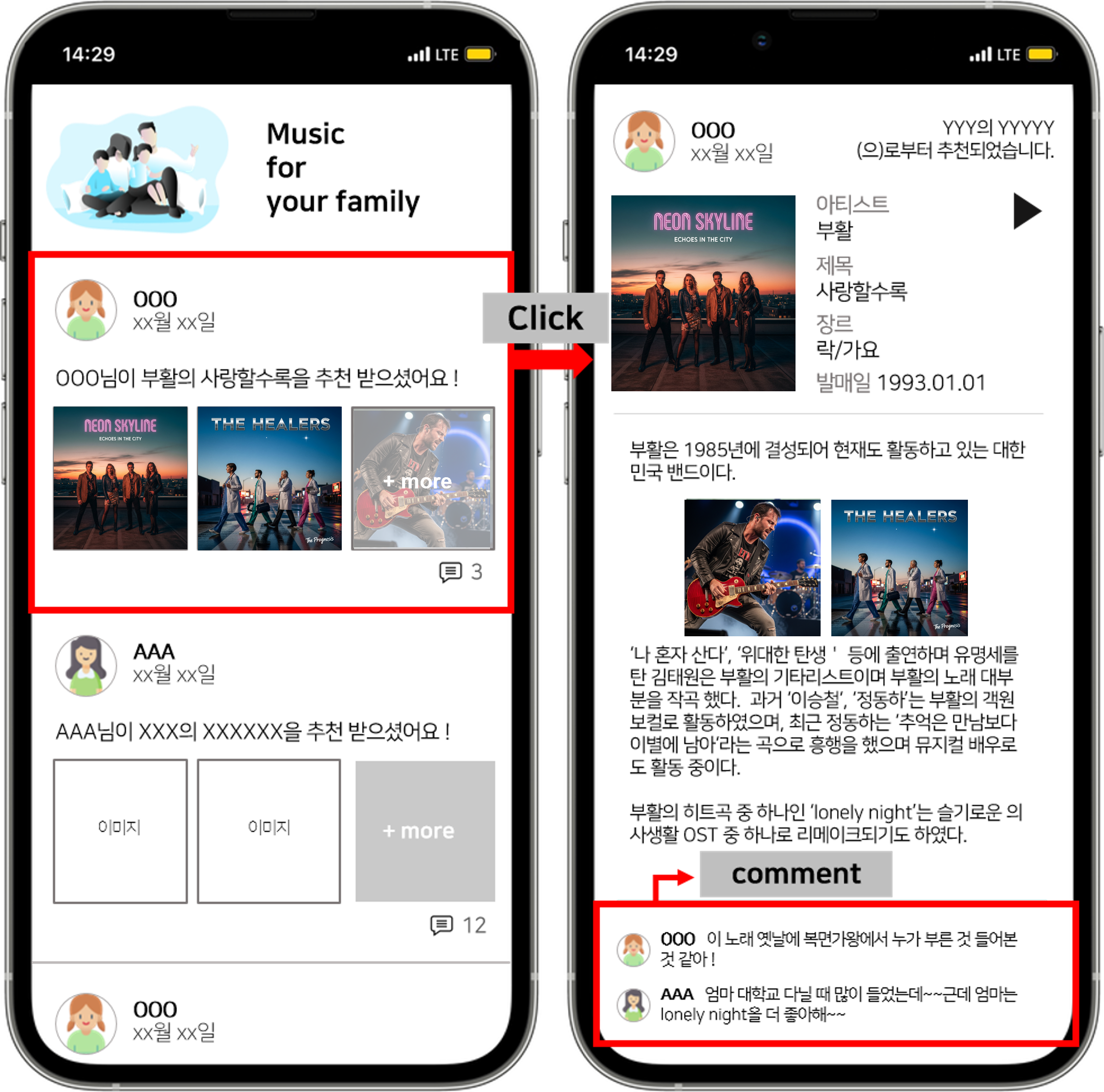}
         \caption{Social media format}
         \label{fig:storyboard_community}
    \end{subfigure}
        \caption{Examples of mock-up app screens. (a): Users chat with their parent/child through a messenger-format app screen. Clicking the shared message of a recommended song, they can browse music-related information. (b): A recommended song is posted on a social media-format app screen. Clicking the post, users can browse music-related information and communicate with their parent/child through comments on the post. The album art shown is AI-generated and used for illustrative purposes only in this publication.}
        \label{fig:Familycom_stoaryboard}
        \hfill
\end{figure}

To explore the ideal mobile app format for music-based communication, we considered two popular options: a messenger format (e.g., WhatsApp) and a social media format (e.g., Facebook). While Messenger apps are widely used by both generations, making the format familiar, the focus on music can get lost as conversations grow, with shared music and discussions easily forgotten. In contrast, social media formats focus more on organizing shared music and related discussions in one place, but they may be less familiar to parents and harder for them to navigate. We asked participants for their preferences between these two formats. To help them visualize the concepts, we presented mock-up screens of each (see Figures~\ref{fig:storyboard_messenger} and~\ref{fig:storyboard_community}).


Most participants (ten out of 13) preferred the messenger format over the social media format. They responded positively about it for reasons such as the convenience for their parents to use and the ease of use for communication purposes: \textit{"A messenger looks more familiar for my parents. They share photos and videos with me using KakaoTalk."} (P\_C9), \textit{"I think a messenger format is better for frequent checks. It will be also easier than comments on social media for my parents to use together."} (P\_C4). Others echoed these sentiments, emphasizing its usability and communication focus: \textit{"Messenger is more convenient. I don't need all the recommended songs gathered in one place."} (P\_C1). 
The remaining three participants favored the social media format, though one expressed reservations: "Social media looks better for browsing all the songs, but it's easier to go back-and-forth (`tiki-taka') in messenger."



\subsubsection{What is the most appropriate way to provide song recommendations and share listening status?}

When comparing on-demand recommendations versus automatic suggestions after finishing a track, most participants preferred the on-demand option: 
\textit{"It might be tiring to get recommendations... if the same song appears repeatedly, I won’t tap it, but if not, I’ll check it more often."} (P\_C8), \textit{"When I get notifications from my music streaming service, it feels like an ad, so I’d prefer to receive recommendations only when I want them."} (P\_C9). However, participants also mentioned that manually tapping a request button every time would be inconvenient. In the end, they agreed that displaying recommendations on the playback screen would be ideal.

For listening status sharing, two options were considered: automatically notifying others when a recommended song is played, versus manually choosing which songs to share. Most participants favored manual selection, citing concerns about privacy and real-time interruptions: \textit{"Real-time notifications could be bothersome"} (P\_C3), \textit{"Automatically sharing what I’m listening to with my parents would feel uncomfortable."} (P\_C4).

\subsubsection{What kind of music-related information would be helpful in facilitating communication through music?}

Since DJ-Fam ultimately aims at creating meaningful conversations between parents and children, we explored various music-related information that could enrich their conversations and evoke memories. We asked participants for their opinions on additional information, and 11 expressed a favorable view, noting that such insights could enhance their discussions and trigger reminiscences: 
\textit{"It would be nice to talk about the lyrics, what my parents did in those days, and what my parents looked like,"} (P\_C8), \textit{"If we have this, the other person doesn't have to explain everything about the song. I think it can provide some knowledge in advance, helping me converse with my parents."} (P\_C12). However, they also commented too many details would not be needed. 

The following information was identified as potentially useful in the interviews: 




\noindent \textbf{Source of a recommended song}: 
Our intuition was that if a recommended song is shared along with the source song that triggered the recommendation, it could spark curiosity and potentially facilitate communication. Most participants agreed, commenting that they would be interested in whether the source music would suit their taste when their parents shared a recommended song: \textit{I would wonder what song my parent listened to that resulted in this recommendation. I’d want to listen to that song, too."} (P\_C10), \textit{"I think it's fun to know it. I can perhaps start a conversation using the source song."} (P\_C14).

\noindent \textbf{Singer's other hit songs}: 
Some participants expressed interest in exploring other songs by artists they like: \textit{"If I like a singer, I can listen to more songs in a similar style."} (P\_C13), \textit{"I prefer recommendations related to specific artists over similar songs, so information on the artist's other hits is valuable."} (P\_C15).



\noindent \textbf{Cover/original songs}: 
Cover songs and their originals might bridge generational gaps, evoking nostalgia for older generations while reflecting current trends for younger ones. Some participants strongly agreed that cover songs could create a shared context for communication and serve as a compelling conversation topic: \textit{"I once listened to a cover song with my parents, and they said they liked it so much; it evoked a sense of nostalgia."} (P\_C11).

%% file: sections/05-2.Implementation.tex
\section{DJ-Fam Prototype Implementation~\label{sec:implementation}}





\begin{figure}
     \centering
     \begin{subfigure}[b]{0.3\textwidth}
         \centering
         \includegraphics[width=0.78\textwidth]{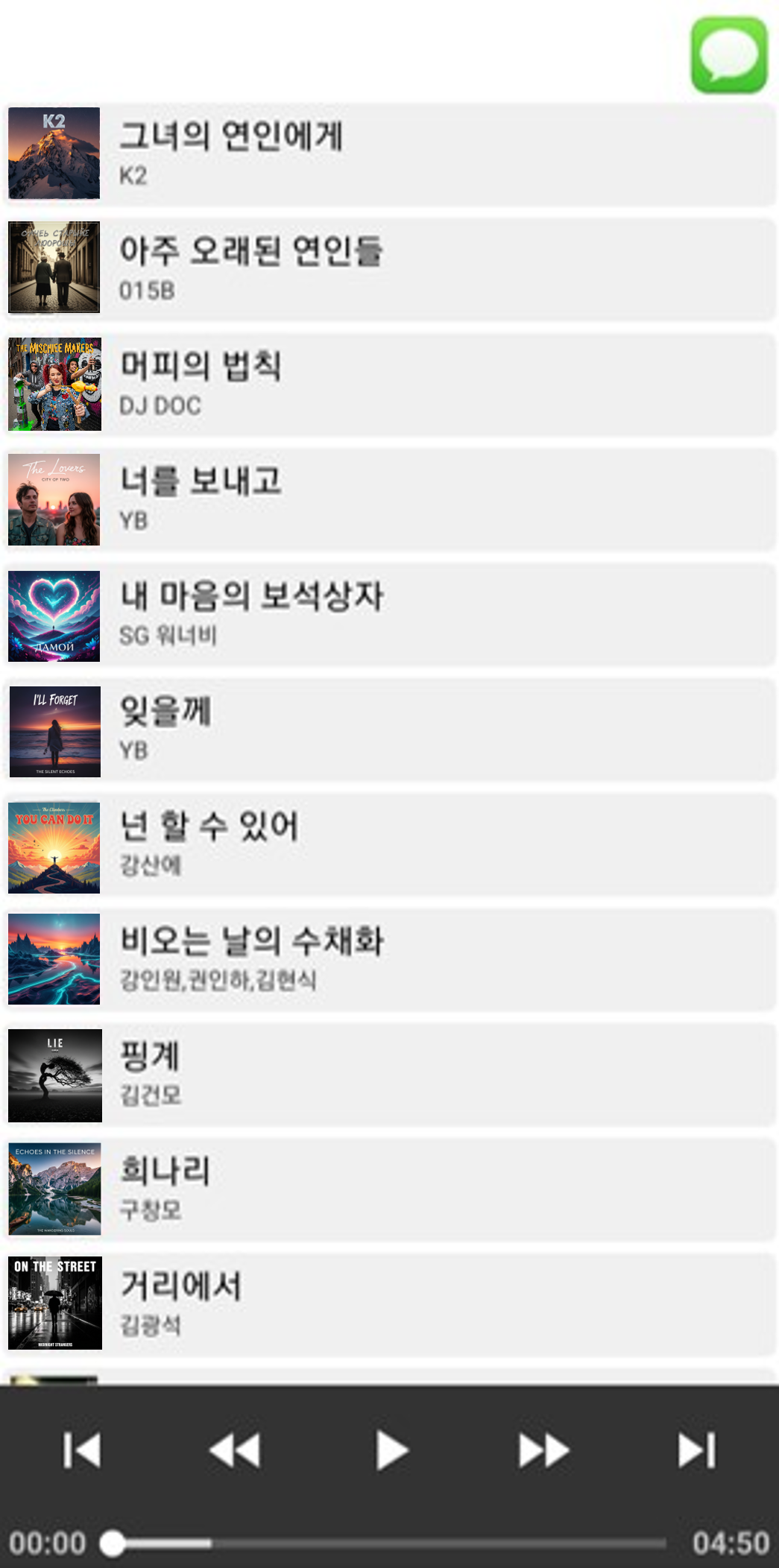}
         \caption{Playlist}
         \label{fig:Playlist}
     \end{subfigure}
     \hfill
     \begin{subfigure}[b]{0.3\textwidth}
         \centering
         \includegraphics[width=0.82\textwidth]{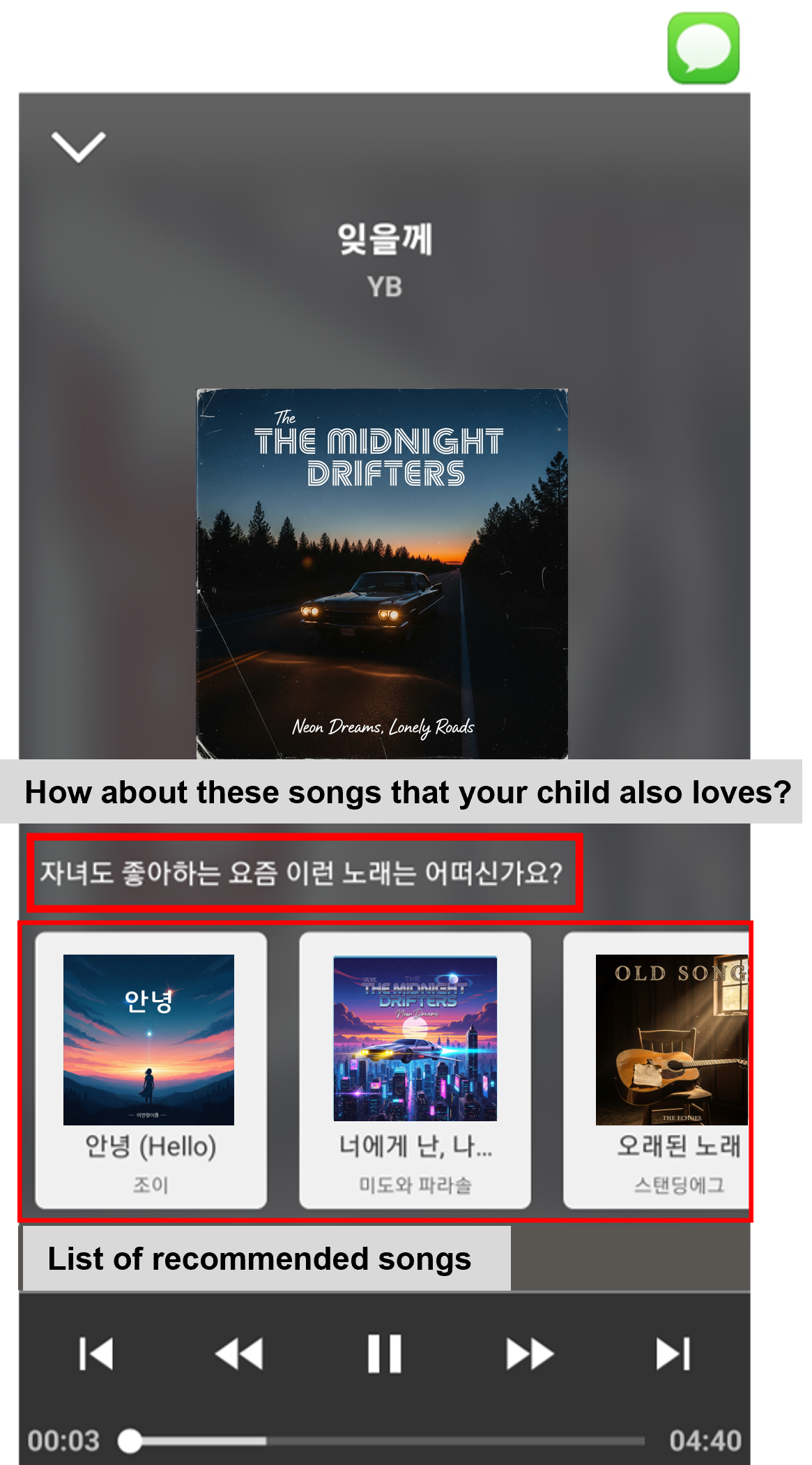}
         \caption{Playing music}
         \label{fig:Playing_music}
     \end{subfigure}
     \hfill
     \begin{subfigure}[b]{0.3\textwidth}
         \centering
         \includegraphics[width=0.79\textwidth]{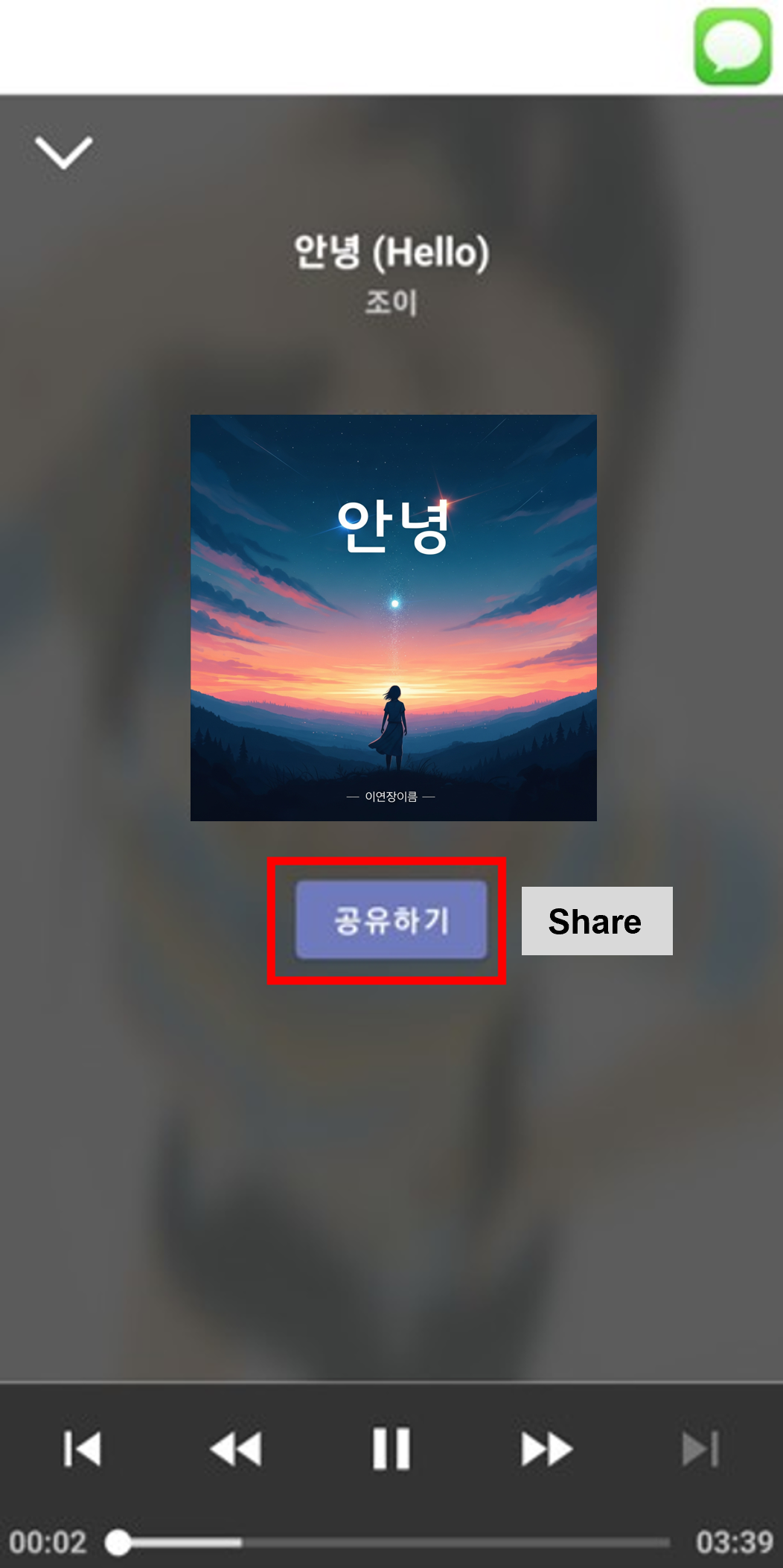}
         \caption{Recommended music}
         \label{fig:Recommended_music}
     \end{subfigure}
     \hfill
     \vspace{0.1in}
     \begin{subfigure}[b]{0.3\textwidth}
         \centering
         \includegraphics[width=0.82\textwidth]{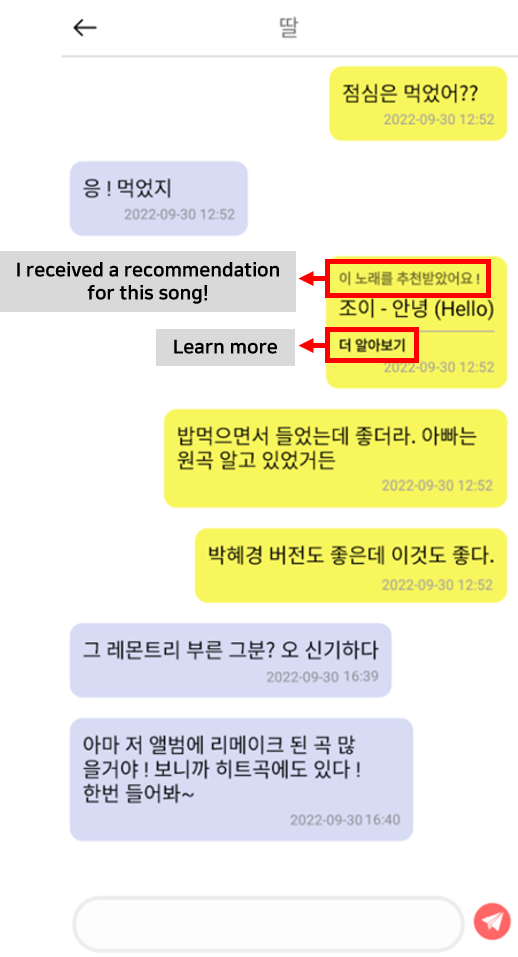}
         \caption{Text messages}
         \label{fig:Message}
    \end{subfigure}
     \hfill
     \begin{subfigure}[b]{0.3\textwidth}
         \centering
         \includegraphics[width=0.82\textwidth]{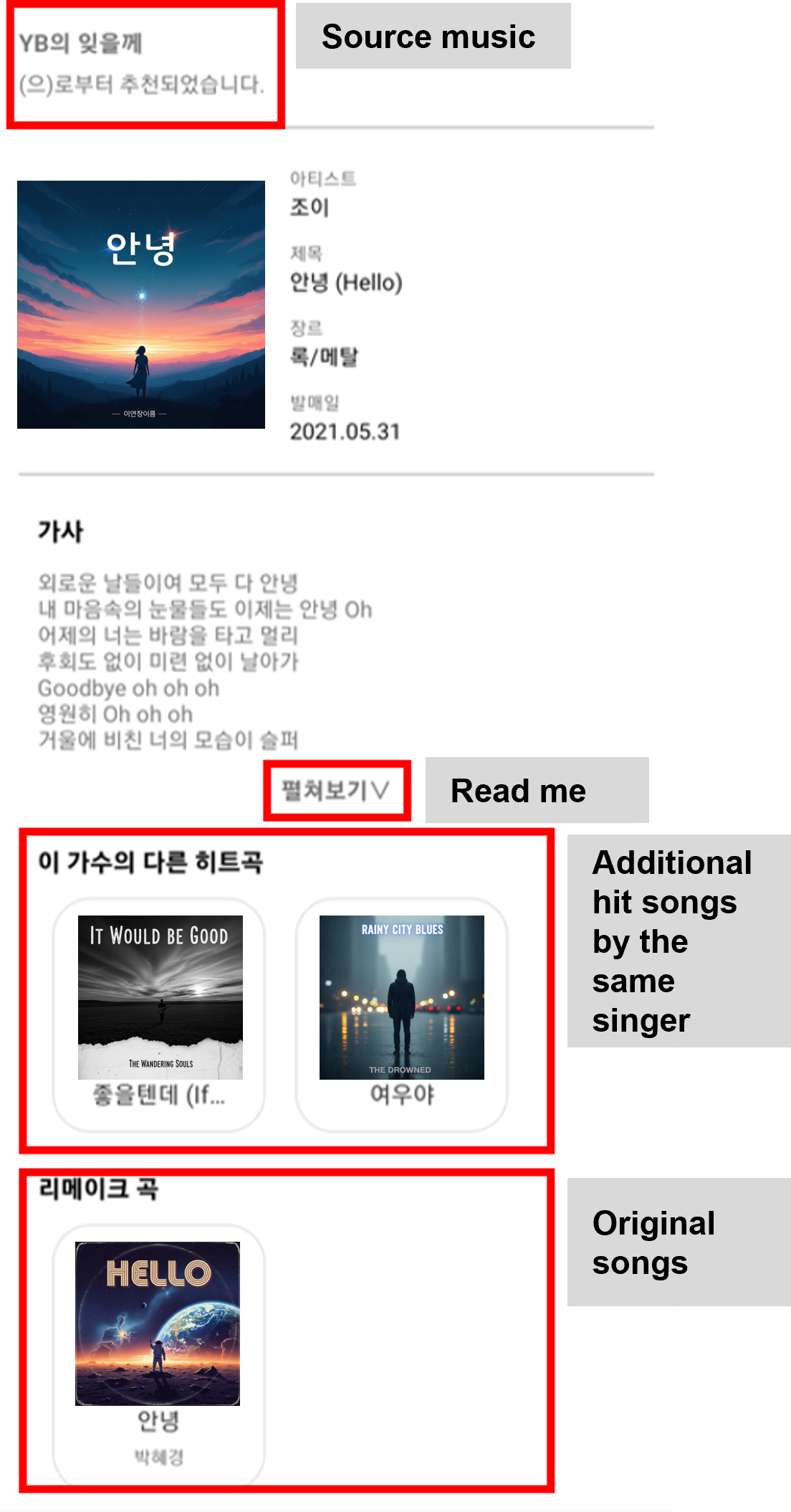}
         \caption{Music information}
         \label{fig:Music_info}
     \end{subfigure}
     \hfill
     \begin{subfigure}[b]{0.3\textwidth}
         \centering
         \includegraphics[width=0.82\textwidth]{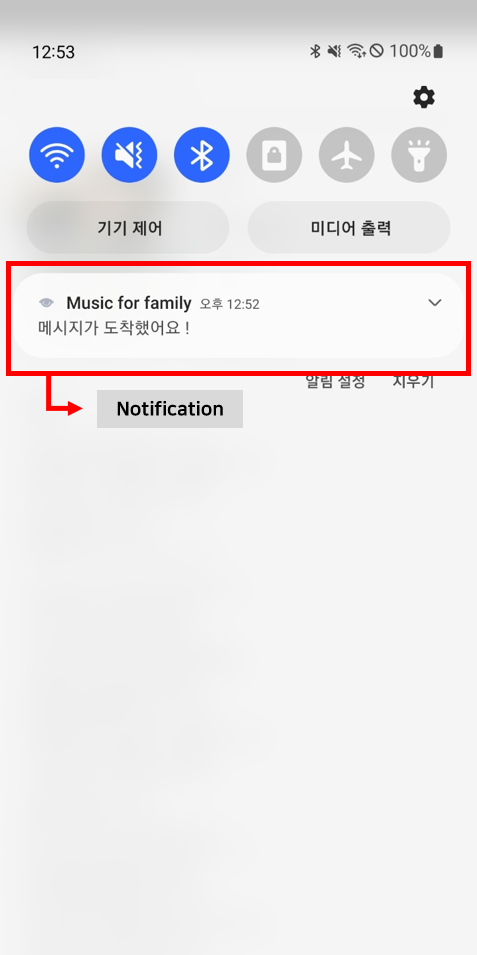}
         \caption{Message notification}
         \label{fig:Message_noti}
     \end{subfigure}
     
        \caption{Example screenshots of the DJ-Fam app interface. All album art shown is AI-generated and used for illustrative purposes only in this publication. The deployed app displayed actual album art to users.}
        \label{fig:Familycom_screenshot}
        \hfill
        \vspace{-0.2in}
\end{figure}

Based on insights from the design study, we developed the DJ-Fam prototype (see Figure~\ref{fig:Familycom_screenshot}). To ensure ease of use, the app’s design mirrors existing music streaming and instant messaging apps. The prototype was implemented on Android, with Firebase\footnote{\url{https://firebase.google.com/}} for messaging and managing databases songs and users.


\subsection{User-side Service Flow}

\textbf{Music playback: }
Upon launching the app, users are presented with a playlist of their frequently played songs (see Figure~\ref{fig:Playlist}). When selecting a song, the playback screen appears (Figure~\ref{fig:Playing_music}), displaying basic information such as the song title, artist, and album art. Beneath the album art, users see the caption "How about these songs that your child (or parent) also loves?", along with recommended songs based on the opposite generation’s playlist—specifically, a child’s playlist for the parent and a parent’s playlist for the child. Tapping a recommended song begins playback, showing the updated screen (Figure~\ref{fig:Recommended_music}). Users can share a recommended song by tapping the share button below the album art.


\noindent \textbf{Music sharing and messaging: } When users tap the share button, they are automatically directed to the messaging screen (Figure~\ref{fig:Message}). A notification is sent to the recipient, informing them of the shared music (Figure~\ref{fig:Message_noti}), inviting them to join the chat. Users can also access the chat at any time by tapping the speech bubble icon in the top right corner of the playback screen. In the chat, users can send messages, and if they shared a song, a special message appears with the caption "I received a recommendation for this song!" along with a 'Learn more' button (Figure~\ref{fig:Message}). 


\noindent \textbf{Music information: }
Tapping the "Learn More" button in the shared song message opens the song information screen (Figure~\ref{fig:Music_info}). This screen provides additional details beyond the basic information like album art, artist, title, genre, release date, and lyrics. It also displays the source song, other hits by the artist, and cover versions, as discussed in Section~\ref{sec:design}. The lyrics are shown in part, with a "Read More" option to view the full lyrics. For popular artists with many hits, only the top three songs are displayed, based on data from a major Korean music streaming service. If the song has cover versions, they are shown; if the song itself is a cover, the original is displayed.


\begin{figure}[t]
    \centering
    \includegraphics[width=0.65\textwidth]{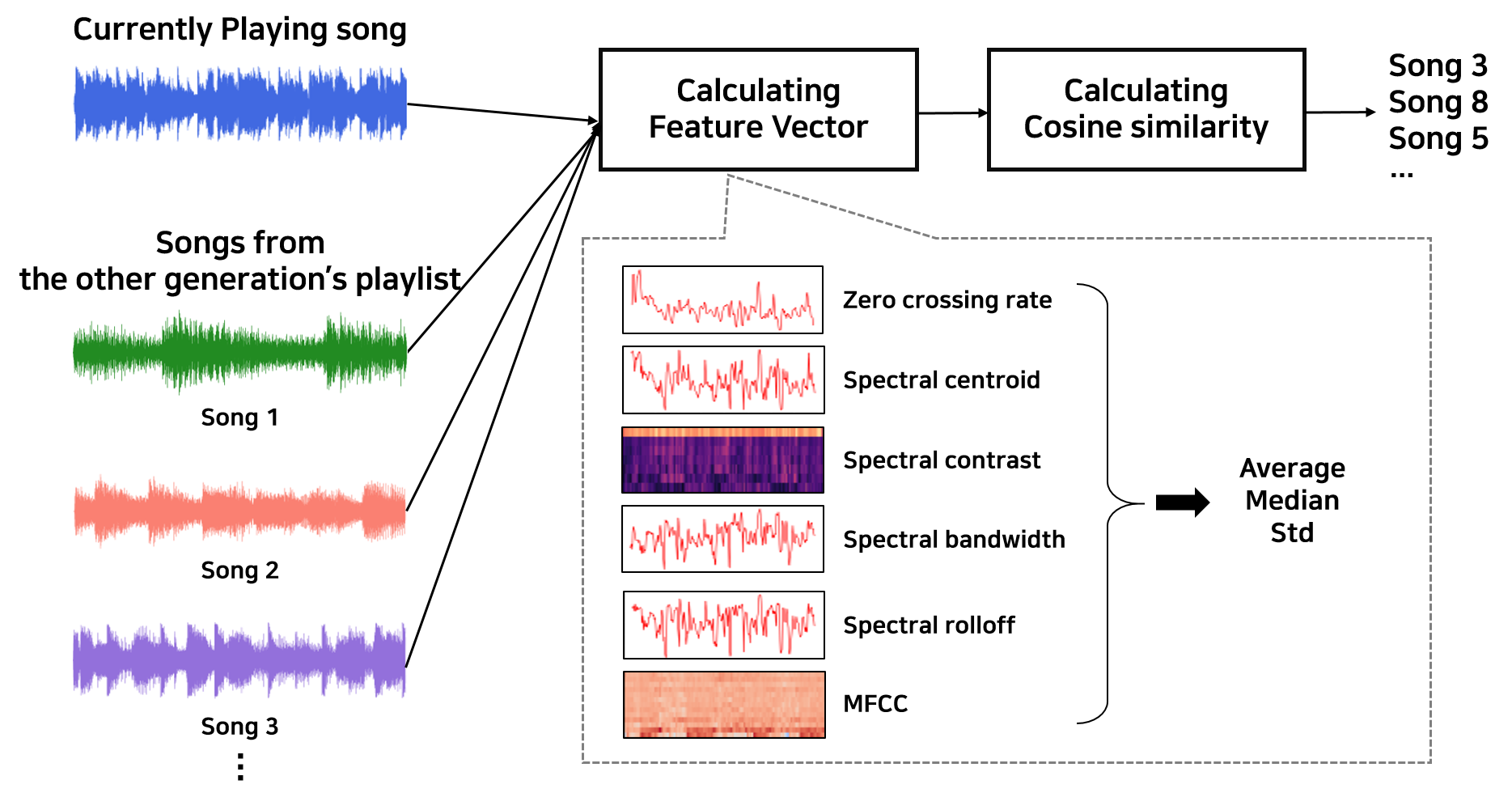}
    \vspace{-0.1in}
    \caption{Process to select songs for recommendation~\label{fig:Music_recommendation}}
    \vspace{-0.15in}
\end{figure}

\subsection{Selecting Songs for Recommendation~\label{subsec:selecting_songs_for_recomendation}} 


The recommended songs selected from the playlists of the other generation should reflect the user's musical preferences to motivate the user to listen to the songs. In this work, we assumed that songs with similar musical features are more likely to match those preferences. Therefore, recommendations are made by comparing the musical features of the currently playing song with songs from the other generation’s playlist. Note that developing and evaluating a method to select songs that precisely match one's preferences is beyond the scope of the current work, which needs further technical development. Here we adopt an existing approach to analyze and compare musical features for fast prototyping and deployment.


Figure~\ref{fig:Music_recommendation} outlines the recommendation process. First, we extract features such as zero crossing rate, spectral centroid, spectral contrast, spectral bandwidth, spectral roll-off, and MFCC from both the currently played song and the songs in the other generation’s playlist~\cite{elbir2018music}. To handle the varying lengths of songs, we calculate the mean, median, and standard deviation for each feature, creating a feature vector. Next, we calculate cosine similarities between the feature vectors of the currently played song and each song in the other generation’s playlist. The top five songs with the highest similarity scores are then recommended. 

%% file: sections/06.Evaluation.tex
\section{Deployment Study }
\label{sec:deployment_study}

In the deployment study, we aimed to examine the overall experiences of participants with DJ-Fam and its effect on parent-child interaction during the four-week usage of DJ-Fam. In the following, we present our findings from the weekly and post-deployment interviews with seven parent-child dyads participating in the study.

\subsection{Extension of Interaction Opportunities through Asynchrony and Music}

\begin{figure}[]
\begin{minipage}[t]{0.43\linewidth}
    \includegraphics[width=\columnwidth]{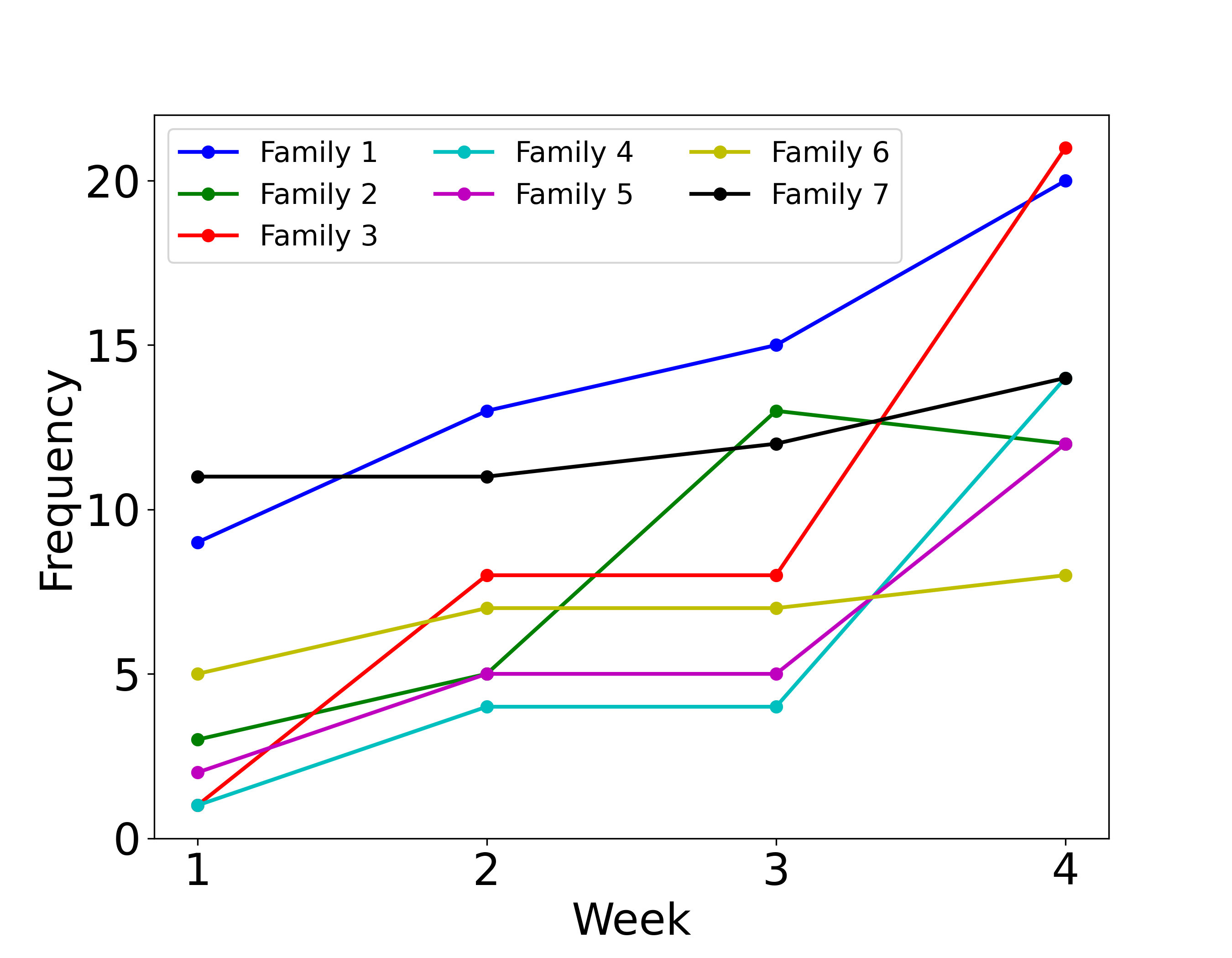}
    \caption{Communication using DJ-Fam.
    \label{fig:Interaction using Familycom per team}}
\end{minipage}%
    \hspace{0.1in}
\begin{minipage}[t]{0.43\linewidth}
        \includegraphics[width=\columnwidth]{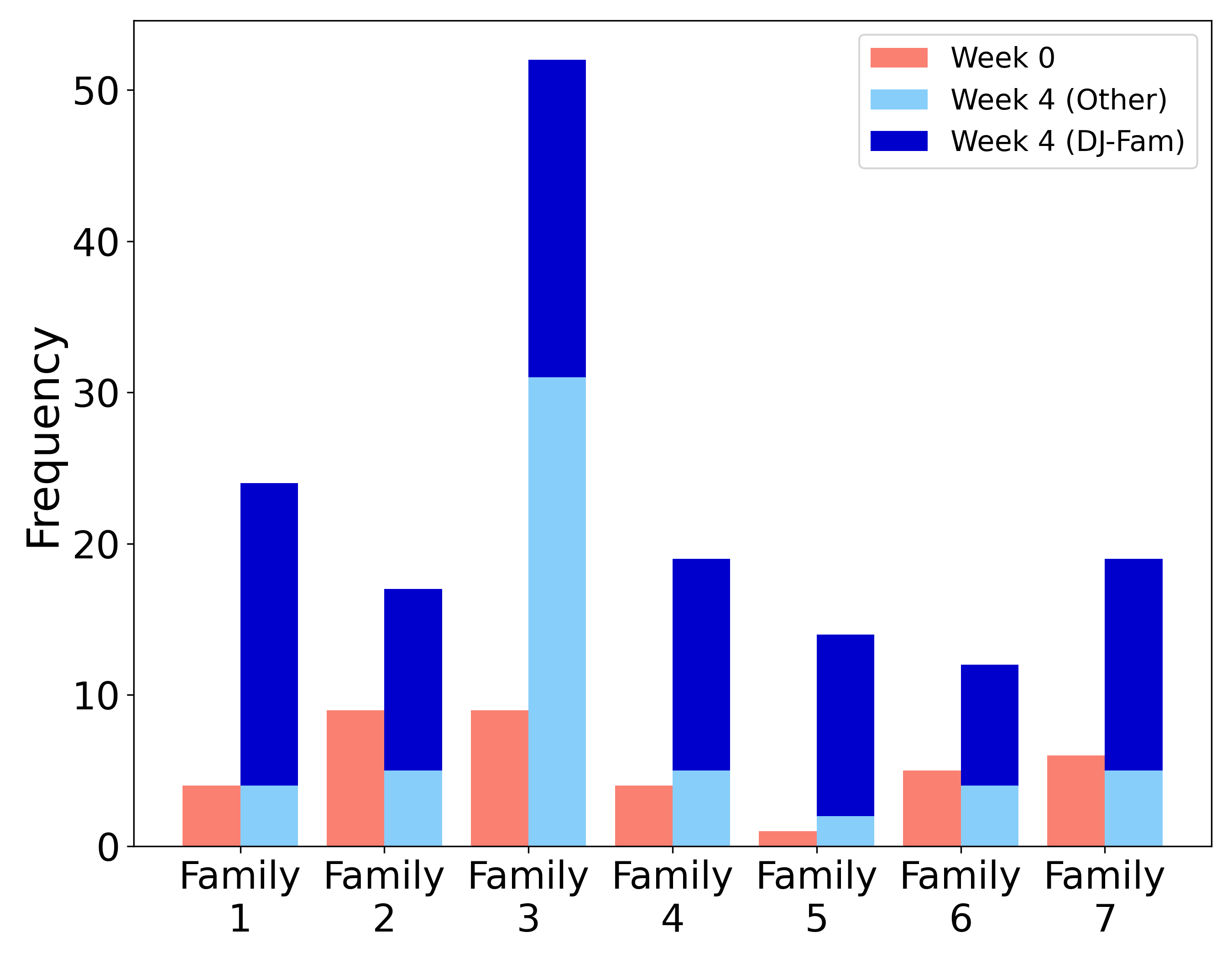}
    \caption{Comparison of before and after the deployment study. \label{fig:Comparison before and after per team}}
\end{minipage} 
\vspace{-0.1in}
\end{figure}

We believe that the two core elements of DJ-Fam - first, using music as an anchor to connect family members, and second, employing asynchronous messaging — serve as an effective means of extending opportunities for family interactions. The frequency of these interactions clearly increased. We first examined the frequency of messaging via DJ-Fam, excluding other forms of communication such as phone calls or other messaging apps. Instead of counting individual messages, we measured \textit{messaging sessions}, where a session is defined as a series of related messages shared between participants to reflect communication frequency better. The participants were asked to count the sessions themselves to preserve privacy. As shown in Figure~\ref{fig:Interaction using Familycom per team}, the number of message sessions increased for all families during the deployment period, with a sharp rise for some, up to 21 times more by the end compared to week 1.

 
We also examined the overall interaction frequency while including other forms of communication. Figure~\ref{fig:Comparison before and after per team} compares the overall frequency between week 0 (just before the study) and week 4, showing a consistent and significant increase across all families. While families increasingly used DJ-Fam throughout the deployment, they maintained similar usage levels for other communication methods. These results suggest that DJ-Fam is creating new interaction opportunities and increasing overall communication frequency, rather than replacing existing channels, by introducing music into conversations where it had not been a topic before. An exception was Family 3, where both DJ-Fam and other methods showed a significant increase, which was later explained by a special family matter during week 4.

We also found that the asynchronous nature of DJ-Fam played a crucial role in its impact. Four out of seven families even reported a shift in their primary communication channel to DJ-Fam.  
This change was positively received, as it provided a new means of communication that offered more opportunities for their interaction:
\textit{"Before, we only talked on the phone, but we can talk more through text messages"} (P\_P3), \textit{"We usually checked in with each other once a week over the phone, but after using this app, we were able to stay in touch even during spare time."} (P\_C1), \textit{"Although phone calls are more convenient, we were able to be free from the constraints of time"} (P\_C2).

Although the asynchrony of DJ-Fam limits the ability to fully share listening moments as in other works, e.g., FamilySong~\cite{tibau2019familysong}, it offers flexibility, allowing participants to engage in interactions at their preferred times throughout their daily routines, such as during their commute, while relaxing at home, or while engaged in other activities. Participants used DJ-Fam in a similar way to how they typically use music streaming and messaging services: \textit{"I used it during my commute, and sometimes I used it in the office."} (P\_P2), \textit{"I often used DJ-Fam at night while coding, and it seemed like I used it a lot while walking."} (P\_C3), \textit{"I usually used it around evening time when I was relaxing at home or lying down. ... And I often used it when I was commuting during weekdays." } (P\_C7).  
Some explicitly noted that it reduces the burden of contacting busy family members while still facilitating more meaningful and frequent communication than phone calls:
\textit{"I feel like there's been much more communication. It was burdensome and difficult to contact my children before because they were always busy, and it felt like a one-sided conversation. However, using this app, we could have better conversations [through messaging], and the burden has decreased."} (P\_P1), \textit{"Generally, phone calls were more about simply checking in or having brief conversations, rather than engaging in long dialogues ... If I want to have a somewhat longer conversation and exchange everyday greetings with someone, then I feel like a message holds more significance." } (P\_P7).




\subsection{Recommendations Creating More Concrete Interaction Opportunities}
DJ-Fam's recommendations provided participants with concrete prompts that could be used immediately to initiate interaction between family members. The recommended songs were consistently listened to when received, with an average of 4.9 songs consumed per week (parents: 5.3, children: 4.5). This figure is a conservative estimate since repeated listening, which participants frequently mentioned by participants, was not counted. There were no significant differences in song consumption between parents and children.
Both parents (an average of 2.8 times per week) and children (an average of 2.9 times per week) consistently shared the recommended songs they listened to through text messages throughout the deployment. Specifically, parent and child participants shared an average of 2.29 and 2.0 recommended songs in the first week, respectively. By the fourth week, they shared 3.86 and 3.71 songs on average, showing a consistent upward trend.




Participants were generally satisfied with the recommended songs. While they often enjoyed listening to them and shared the songs they listened to, there were generational differences in the reasons for doing so. Upon receiving recommendations from the parent's playlist, the children tended to select songs by familiar artists or those with recognizable titles, making their choices more aligned with their preferences. In particular, P\_C5 said, \textit{"I listened to it because it was a song by a singer I used to like, and I was happy because it was a song I knew. `Right, there was such a song like this.' I thought."} P\_C2 said, \textit{"Even the recommended songs were good because it vibes similar to the songs I chose [for my playlist]."} The child participants shared the songs to talk about themselves, such as unfamiliar songs of musicians they know, originals of famous cover songs, or their musical tastes: \textit{"I want to tell my feeling about the songs."} (P\_C2), \textit{"I want to express that I know this singer and tell my taste." } (P\_C4). 

On the other hand, parents were more open to new songs recommended from their child's playlist. They often enjoyed these songs even when hearing them for the first time and embraced unfamiliar recommendations that appealed to them or sparked their curiosity. Parents also expressed empathy toward their children's musical preferences often. When sharing the recommended songs, they sought to connect with their children by sharing personal anecdotes from their youth—such as experiences from their 20s or 30s—and reminiscing about music-related memories. They also used recommendations to suggest spending time together in the future: \textit{"I shared the song because it brought back memories of listening to it with my children and resonated with my emotions."} (P\_P4), \textit{"I wanted to share it because it's a song my son and I used to listen to together. I wanted to say something like, `Remember this? We used to listen to it a lot when you were young, or maybe we used to hum it all the time.'"} (P\_P6).

\subsection{Beyond Sharing Songs: Fostering Intimacy through Broader Conversations}
Despite the asynchronous and sometimes interleaved nature of the interactions, using music as the conversation theme and providing tailored song recommendations frequently sparked more intimate family discussions, covering a wider range of topics. Eleven participants reported that they moved beyond everyday, mundane conversation topics and engaged in richer, more novel discussions. Three other participants noted that while their conversation topics remained similar, they were able to discuss the songs as an additional topic. Additionally, three parents mentioned that their conversations became more prolonged compared to before. Moreover, six participants (four parents and two children) reported they felt changes in intimacy with their child or parent. They mentioned that engaging in more conversations not only about music but also about their daily lives led to deeper dialogues and more closeness between generations.

\subsubsection{Song recommendations as conversation openers}
Song recommendations often served as conversation starters, beginning with discussions about participants' feelings toward the songs and frequently expanding to conversations about other songs by the same artist, their musical preferences, or even leading to new song recommendations. For instance, when children shared recommended songs via text, most parents would commonly ask how their child had discovered the songs. The children would then explain how they knew the songs or the artists, naturally leading to discussions about the singer's background, TV shows featuring the songs, cover versions, or original recordings. For example, one participant shared: \textit{"I thought Singer A was an actor, but my mom told me he was originally a singer... when I asked about a singer unfamiliar to me, she mentioned a duet song that the singer had performed with a famous entertainer."} (P\_C1), \textit{"When I listened to the song, I remembered a TV show where it was played, so I shared it to talk about the show." } (P\_C5). Similarly, parents would share stories when recommending songs to their children. One parent explained, \textit{"I told my child that I had liked the original version of this cover song in the past."} (P\_P4) Another participant recalled, \textit{"The song my father shared with me was a cover version. He asked how I knew it, and I replied that it was part of a TV series soundtrack. Then he told me, `The singer of the original song is ... you probably didn't know that.'" } (P\_C3).

\subsubsection{Sharing experiences related to the songs} 
Conversations often evolved into discussions about experiences related to the songs, such as stories from when the songs were released or other personal memories tied to them. These exchanges evoked nostalgia for the parents and sparked curiosity and interest in the children. One parent shared: \textit{"At that time, there were music festivals, and singers performed there... There aren't many singers like that now, but listening to those songs helps me share forgotten stories with my child and brings back memories."} (P\_P1), \textit{"My mom told me that while listening to the songs, she remembered the people she met in those days and missed those times. It was interesting." } (P\_C5). 

Parents also shared experiences they had with their children. When a recommended song was one their child had introduced to them in the past or a song they had listened to together, it often led to meaningful conversations. For example, one child said, \textit{"One of the recommended songs was one I had often mentioned to my dad before. So when he saw the recommended song, he sent me a message saying, `Here's the song you talked about before.'"} (P\_C3), \textit{"I remembered hearing that song in a drama and singing it at karaoke, and I talked about it with my son." } (P\_P6). One child recalled, \textit{"My mom shared a song we had listened to together while traveling and said, `Let's go on a trip together again,' reminiscing about our past trip. She also sent me a long message and said she felt like crying. Having these conversations, which we hadn't had before, felt unfamiliar, but it was nice." } (P\_C5).

\subsubsection{More frequent and richer updates, and family activity planning} 
The conversations triggered by music often transitioned into daily topics, including questions such as `Have you eaten?', `What are you doing?' or `Where are you?' Participants reported a seamless transition from discussions about music-related topics to everyday life conversations, resulting in heightened frequency and duration of interaction: \textit{"I didn't talk much about everyday things with my parents, but while using this app, I naturally started having conversations about daily life, not just about songs."} (P\_C1), 
\textit{"We also talked about daily life ... As we talked about music-related topics, it naturally led to conversations like `Have you eaten?' or `What did you order for food?' and `How's your cold?'"} (P\_P7). Most participants expressed great satisfaction with the natural flow of conversation transitioning from music to everyday topics: \textit{"In the past when I talked to my daughter, we usually just asked after each other briefly. But now, being able to have longer conversations about daily life has increased our intimacy, which I appreciate."} (P\_P3), \textit{"Anyway, having music as a conversation medium, while doing that, we could chat with each other, and also, during that time, we talked about our daily lives, so it was nice." (P\_C7).}

Meanwhile, aside from conversations in DJ-Fam's message channel, various forms of interaction were also triggered.
For example, some participants even discussed music on other communication channels, such as phone calls or face-to-face conversations: \textit{"We met a while ago and talked about various things related to a specific singer. While driving, we listened to songs from a shared playlist and discussed why that singer didn't appear."} (P\_P7), \textit{"My mom was watching TV, and she seemed to have discovered a new band. She mentioned that their songs were good and told about what the band does. Generally, she wouldn't have brought up such topics, but she mentioned it while we were on the phone. So, I found that quite interesting." } (P\_C6). Surprisingly, some of them also have planned new activities or events within the family: \textit{"When I sent the song saying it was good, my dad mentioned that it was a song he used to sing with his friends at the karaoke. Later, he suggested we go to the karaoke together and sing it."} (P\_C3), \textit{"We were talking about various things like, `It seemed like people were traveling a lot these days.' and we started discussing the idea of going on a trip together" } (P\_P6).


\subsection{Understanding and Expansion of Musical Tastes}

DJ-Fam facilitates musical exploration between generations, providing participants with a unique experience. The participants responded that using DJ-Fam became an opportunity to understand the other generation's musical tastes. Most parents mentioned that they were unfamiliar with their children's music preferences before the deployment study. After the study, however, they expressed satisfaction with discovering their children's music preferences through recommended songs: \textit{"I can say that the preferences I know of my daughter were based on her childhood. So, I actually didn't know what songs she currently likes and listens to often, but through this opportunity, I realized her current preferences ..."} (P\_P4), \textit{"I liked all the songs my daughter listens to. I felt that our tastes are similar"} (P\_P3).


Surprisingly, while all child participants mentioned they already knew their parent's music preferences well before the deployment study, three responded that they had misunderstood their parent's preferences: \textit{"Contrary to what I thought, my mom kept up with the trend. She liked hip music and trendy things..."} (P\_C5). The other four felt that their parent's tastes were similar to what they already knew. Two mentioned that they felt their parent's tastes they already knew, allowing them to confirm: \textit{"It's just like my father. I realized that my preferences are really identical to my father's."} (P\_C3), \textit{"I realized that my mom's taste hadn't changed. While listening, I felt like our tastes are now similar, and as I grew up listening to the songs my mom often played, I also wondered if I liked this genre of music because of that." } (P\_C7).

Some participants reported that using DJ-Fam facilitated the discovery of their parent/child preferences and influenced the expansion of their musical tastes. Particularly, some parents who had limited knowledge about songs from the young generation tended to enjoy the recommended songs after experiencing them: \textit{"Since we lived separate lives, I didn't have the experiences listening to the favorite songs of my child, so I thought they didn't match my preferences … But now I realize that not all of today's songs are overwhelming or make me feel uncomfortable. I've discovered that songs with the vibe I like still exist …"} (P\_P1), \textit{"I liked the recommended songs. I thought that I would try to listen to the songs of my daughter's generation." } (P\_P4). Interestingly, one parent (P\_P2) said that some songs did not match her taste, but she gradually started to like them after listening to them several times. 

Some children had not enjoyed or were unfamiliar with some songs from their parents' playlist that had an old-fashioned feel. However, a few listened to the songs until the end and eventually showed interest: \textit{"It was better [than I expected]. It was my first time listening from start to finish, and it felt different and good."} (P\_C3), \textit{"I don't usually like slow songs, but [after listening to my parents' songs,] I thought some songs were good. I even shared the songs to express that I like this genre." } (P\_C4).


\subsection{Feedback and Improvement Opportunity~\label{subsec:application_usability}} 
We gathered diverse feedback through post-deployment interviews, with several insights offering direct opportunities for improvement. Below, we elaborate on themes related to UI effectiveness, context of recommendations, and collaborative playlist creation.

\subsubsection{UI effectiveness}
Participants were generally positive about the app’s usability. They noted that they became comfortable with the app quickly, even without prior experience using a combined music playback and messaging app (\textit{e.g., "It wasn't difficult to use ... I can communicate instantly. It was nice to have a messenger integrated." } (P\_P3)). The similarity of the style and interface to the apps they already use was explicitly cited in one of the responses (P\_C1).

While the recommendation feature was viewed as intuitive and engaging, there is potential to further increase its visibility. Currently, recommended songs appear on the playback screen, which participants may miss when they aren’t actively using the app — for instance, while driving, studying, or multitasking with other apps (e.g., shopping). Some participants mentioned the difficulty of checking recommendations while occupied. Improved methods for displaying recommended songs, such as showing pending recommendations when the app regains focus or using a small Picture-in-Picture (PiP) window when other apps are in use, could enhance visibility and ultimately help turn more listening moments into valuable opportunities for parent-child interaction. 

\subsubsection{Context of recommendations}

As mentioned in Section~\ref{sec:design}, DJ-Fam provides additional information about recommended songs to enrich conversation topics, aiming to supplement users' knowledge of the songs and aid in recalling them. However, we found that participants did not engage with this information as much as anticipated. Instead, they preferred discussing what they knew about the songs and sharing personal thoughts and experiences. While some participants used the additional information to browse other popular songs by the artist or to read lyrics, they seldom incorporated it into their conversations. 

However, the \textit{context} of the recommendations emerged as a valuable feature for some participants, particularly the source of each recommended song. Users compared the musical tastes or moods of the recommended songs to those in the recommending generation’s playlist. One family who actively used the source information commented, \textit{"The music information screen let me know what songs I listened to and why these songs were recommended, so I wanted to use this to tell [my son] about things I experienced in the past." } (P\_P1). Another participant shared, \textit{"My mom used this feature first. When she shared a song, she said, `This song is good,’ I wondered about it – it turned out to be from her playlist. We used that context to recommend songs to each other."} (P\_C1). Additionally, one participant suggested knowing how frequently their parent had recently listened to a recommended song would be useful.

As prior studies indicate, knowing the context behind recommendations (e.g., why a song is recommended) is essential for users to engage with it meaningfully~\cite{lee2019can}. We believe that adding more contextual details could increase users' interest in the recommendations and support song selection. For instance, if DJ-Fam highlights that one of the recommended songs was recently enjoyed by the parent, the child may feel more motivated to listen, engaging with the parent's music taste. Over time, more frequent listening to songs from the other generation's playlist can naturally lead to greater chances of parent-child interaction. 

\subsubsection{Collaborative playlist creation}

As DJ-Fam was an experimental prototype, participants offered many suggestions for additional functions, such as enabling music ratings, displaying lyrics in sync with playback, and playback functionality for songs on the music information screen. The most frequent suggestion was about recommendations, mainly asking for a wider pool of tracks (\textit{e.g., "I think I did not listen to diverse recommended songs since there were many similar songs."}, (P\_C5)). Three participants expressed interest in selecting songs for recommendation themselves: \textit{"After seeing the songs my mom liked and the ones we shared, I had some songs I wanted to recommend, which were not recommended by the app."} (P\_C1).

While expanding the pool of songs is an obvious improvement, we observed an interesting pattern during the deployment; as users added recommended songs to their playlists, the overlap between parent and child tracks gradually increased. Over time, this convergence could create a unique playlist featuring a balanced mix of each generation's favorites. This shared playlist may foster empathy and spark conversation between parents and children. A key feature we could introduce is the automatic generation of a collaborative playlist designed by extracting overlapping songs from both playlists. The collaborative playlist could offer users a novel social experience centered around music, e.g., when a family goes for a drive together, they could enjoy a playlist of mutually appreciated songs along the way.

%% file: sections/07.Discussion.tex
\section{Discussion}

\subsection{Trade-off between Co-listening and Interaction Opportunities}

The potential of music to strengthen bonds between remote loved ones has been explored in various ways. A body of works involves creating virtual spaces for sharing music or sounds, which fosters a sense of togetherness through shared sensory experiences. As shown in the prior work~\cite{tibau2019familysong}, this approach is particularly effective for building ongoing connectedness and awareness, as it is built upon temporally synchronized listening moments. On the other hand, asynchronous interaction is a primary design choice that underpins DJ-Fam. While it trades off the ability to share real-time listening moments, it expands opportunities for more intimate interactions. Asynchrony offers great flexibility for coordinating communication in remote settings and helps alleviate privacy concerns, which have been a significant challenge in creating shared spaces.

This strength of asynchrony is evident in the frequent and spontaneous use of DJ-Fam by families. While our study focused on building a concrete case study, broader design directions leveraging asynchrony can be explored by aligning DJ-Fam with prior research on intergenerational communication. Prior work highlights the role of phatic communication—lightweight, low-effort exchanges that maintain relationships— in fostering family bonding \cite{abel2021social,taipale2018big}. An effective approach in this context is to design interactions that integrate seamlessly into existing routines and evolve into family rituals \cite{fiese2002review}. DJ-Fam achieves this by embedding communication opportunities into everyday music-listening moments, an activity already present in many people’s daily lives. Similar approach is also observed in prior work that integrates social engagement into other everyday activities, such as exercising, where features like shared progress tracking, friendly competition \cite{gui2017fitness}, or gaming \cite{park2012exerlink} to enhance motivation and connectedness.

Additionally, this approach resonates with the expectations of younger generations, who sometimes feel burdened by structured or obligatory communication with family~\cite{yarosh2011mediated}. By shifting from calls or interactive messaging to an implicit and expressive medium, DJ-Fam lowers the pressure of interaction while maintaining a sense of presence. At the same time, our study reveals key elements that appeal to the parent generation. For older adults, creating opportunities for intergenerational reminiscence—where shared moments of nostalgia through music serve as a bridge between generations—can foster emotional bonds and enhance psychological well-being \cite{gaggioli2014intergenerational}. Additionally, broadening the scope of intergenerational dialogue beyond routine check-ins through shared cultural artifacts, such as music, can encourage deeper conversations and mutual understanding~\cite{wallace2018associations}. These elements ensure that communication tools are not only facilitating connection but also enriching the quality of interaction in meaningful ways.

\subsection{Relevance and Sustained Engagement in Family Music Recommendations}

Music sharing has been shown to facilitate social interaction by triggering conversations and encouraging the exchange of personal stories~\cite{leong2013revisiting, park2022cross}. Prior research on mobile music sharing~\cite{bassoli2006tuna, hakansson2007facilitating, kirk2016understanding} has predominantly focused on \textit{user-initiated music sharing}, where individuals actively select songs to share—either as a form of self-expression or to foster engagement with the recipient’s tastes~\cite{park2022cross}.  In contrast, DJ-Fam introduces a \textit{system-driven recommendation model} that curates music from the other generation’s playlist. As discussed earlier, this approach alleviates the effort required from users while leveraging existing family bonds to facilitate engagement~\cite{lee2019can}. 

However, a key trade-off emerges between personalization and intergenerational engagement. Participants occasionally expressed reluctance toward certain recommendations due to stark differences in musical taste, which led to skipping songs and, in turn, reduced opportunities for interaction. This highlights the importance of balancing recommendation accuracy with the potential for meaningful family interactions. While these factors may appear to be at odds, their relationship is likely more nuanced. Even when a song is not immediately relevant to the listener’s taste, engagement features such as comments, reactions, listening history, or personalized explanations for why a song was recommended could provide additional context, fostering curiosity and interaction. Exploring mechanisms that make recommendations more conversational—rather than merely algorithmic—could enhance their role as social catalysts.

Given that DJ-Fam operates within the context of family relationships, it is crucial to consider the evolving role of recommendations over time. Our observations revealed certain limitations in the current pool of recommended songs—some participants noted that repeated recommendations or a lack of variety could diminish engagement. These observations led us to consider broader, long-term challenges in sustaining meaningful interaction. Music preferences naturally shift, and to maintain engagement, the system must adapt by refreshing the pool of recommended songs to reflect these changes. Additionally, the nature of intergenerational interaction itself may transform over prolonged use. For instance, the initial novelty of reminiscing over past music may diminish, necessitating new forms of engagement. A longitudinal perspective is essential in defining success metrics for such recommendations—moving beyond immediate listening rates to assessing sustained participation and evolving patterns of family communication.

\subsection{Designing Music-Mediated Nudges for Families}

Behavior change theories \cite{leonard2008richard} have been widely applied in the design of communication nudges to strengthen connectedness among remote family members. By viewing DJ-Fam as a communication nudge, we can extract key design lessons for fostering engagement through choice architecture and contextual triggers.

A primary implication from our deployment study is the framing of recommendations within a structured choice architecture. Rather than presenting an open-ended selection of songs, DJ-Fam curates a targeted set of music choices originating from a family member's expression, making them more emotionally salient. Framing recommendations as personal gestures rather than generic system-generated suggestions increases perceived social value, making users more likely to engage \cite{park2021social}. In our study, this led to rich interactions, ranging from casual daily conversations to emotional self-expression and reminiscence sharing. Similar approaches could be extended to other media, such as photos, video clips, or news articles, where recommendations are framed as personalized social cues rather than pure algorithmic outputs.

Beyond curating song choices, DJ-Fam’s listening status sharing feature functions as a well-timed trigger that leverages reciprocity and social awareness. By notifying users when a family member has engaged with a recommended song, the feature creates a shared context for spontaneous interaction. This aligns with research on ambient awareness \cite{fogg2009behavior}, which emphasizes the role of low-effort social cues in triggering meaningful engagement. While various design approaches exist for fostering such awareness, DJ-Fam employs an explicit, user-initiated action—pressing a sharing button. This small, lightweight gesture reduces the barrier to initiating communication while maintaining privacy control, ensuring that sharing remains voluntary. Additionally, the presence of reciprocal engagement mechanisms introduces a subtle social expectation to reciprocate, a known driver of sustained behavior change \cite{caraban201923}. In our study, parents frequently listened to songs recommended by their children, reinforcing a feedback loop of mutual engagement. This mirrors prior findings on reciprocity in social interactions \cite{gamberini2007embedded} and highlights the importance of designing for lightweight but reinforcing communication loops in inter-generational interactions. From a broader perspective, our study identifies a design space for building shared contexts in music listening and communication nudging. Future work could explore different modes of listening status sharing, such as automatic vs. manual triggers, or introduce lightweight engagement features (e.g., likes, emojis) or contextual prompts.

%% file: sections/08.Limitations.tex
\subsection{Cultural \& Demographic Limitations and Opportunities for Generalization}

Our study is inherently shaped by demographic and cultural factors, as the deployment involved seven parent-child dyads, all from South Korea, with parents mostly in their 50s and children in their 20s. This context confines the scope of our findings, as family dynamics, communication styles, and engagement with music can vary across cultures and generations.

Nevertheless, the core design principles of DJ-Fam—leveraging music as a communication medium and incorporating lightweight interaction—are not inherently culture-specific. With appropriate adaptations to different cultural and family contexts, similar approaches could be applied in a broader range of settings. At a fundamental level, DJ-Fam relies on widely used technologies such as mobile devices, music streaming services, and messaging applications, minimizing technical barriers to broader adoption. As long as these underlying technologies remain popular, the potential user base of such systems is likely to grow over time, further increasing opportunities for generalization.

That said, the effectiveness of music as a medium for parent-child interaction may differ across cultural contexts. According to the IFPI Global Music Report 2025~\footnote{\url{https://www.ifpi.org/wp-content/uploads/2024/03/GMR2025_SOTI.pdf}}, South Korea ranks 7th globally in music market development, indicating a strong cultural emphasis on music consumption in daily life. It suggests that the extent to which music fosters social interaction and strengthens relationships may be culturally contingent. Prior research also highlights cross-cultural differences in the role of music, with variations in how music influences social bonding and emotional expression~\cite{boer2012towards, boer2012young}. For example, these studies describe how music serves different social functions in individualistic and collectivistic cultures, which could inform recommendation strategies by tailoring the balance between personal expression and shared musical experiences.

Beyond cultural factors, generalization strategies should account for variations in family communication patterns, shaped by cultural norms, individual personalities, and values. Family Communication Patterns Theory~\cite{koerner2006family} categorizes families into conversation-oriented, who prioritize open dialogue and mutual respect, and conformity-oriented, who emphasize unity and adherence to shared values. Such differences could be reflected in recommendation strategies; for instance, systems could adapt by either diversifying musical exposure or reinforcing shared tastes, depending on the family's communication style. Tailoring these aspects could enhance the system’s relevance across diverse family structures and cultural settings.

%% file: sections/09.Conclusion.tex
\section{Conclusion}

This paper presents DJ-Fam, a mobile application that utilizes music-listening moments to foster empathy and create opportunities for meaningful conversations between parents and young adult children living apart. To achieve this goal, DJ-Fam features two main design components: recommending songs from the other generation's playlist and sharing the status of listening to recommended songs. It allows parents and children to listen to and share the other generation's favorite songs, which provides a shared context for engaging conversations. Results from the four-week deployment study with seven parent-child dyads in South Korea demonstrated that DJ-Fam has the potential to foster enriched parent-child interaction. We also present a discussion of design implications, including opportunities to promote interaction, insights for designing communication nudges, considerations for music recommendation within family contexts, and reflections on cultural and demographic limitations, along with strategies for broader applicability.

%% file: main.bbl

\begin{thebibliography}{88}


\ifx \showCODEN    \undefined \def \showCODEN     #1{\unskip}     \fi
\ifx \showDOI      \undefined \def \showDOI       #1{#1}\fi
\ifx \showISBNx    \undefined \def \showISBNx     #1{\unskip}     \fi
\ifx \showISBNxiii \undefined \def \showISBNxiii  #1{\unskip}     \fi
\ifx \showISSN     \undefined \def \showISSN      #1{\unskip}     \fi
\ifx \showLCCN     \undefined \def \showLCCN      #1{\unskip}     \fi
\ifx \shownote     \undefined \def \shownote      #1{#1}          \fi
\ifx \showarticletitle \undefined \def \showarticletitle #1{#1}   \fi
\ifx \showURL      \undefined \def \showURL       {\relax}        \fi
\providecommand\bibfield[2]{#2}
\providecommand\bibinfo[2]{#2}
\providecommand\natexlab[1]{#1}
\providecommand\showeprint[2][]{arXiv:#2}

\bibitem[Abel et~al\mbox{.}(2021)]%
        {abel2021social}
\bibfield{author}{\bibinfo{person}{Susan Abel}, \bibinfo{person}{Tanya Machin}, {and} \bibinfo{person}{Charlotte Brownlow}.} \bibinfo{year}{2021}\natexlab{}.
\newblock \showarticletitle{Social media, rituals, and long-distance family relationship maintenance: A mixed-methods systematic review}.
\newblock \bibinfo{journal}{\emph{New Media \& Society}} \bibinfo{volume}{23}, \bibinfo{number}{3} (\bibinfo{year}{2021}), \bibinfo{pages}{632--654}.
\newblock


\bibitem[Arnett(2000)]%
        {arnett2000emerging}
\bibfield{author}{\bibinfo{person}{Jeffrey~Jensen Arnett}.} \bibinfo{year}{2000}\natexlab{}.
\newblock \showarticletitle{Emerging adulthood: A theory of development from the late teens through the twenties.}
\newblock \bibinfo{journal}{\emph{American psychologist}} \bibinfo{volume}{55}, \bibinfo{number}{5} (\bibinfo{year}{2000}), \bibinfo{pages}{469}.
\newblock


\bibitem[Arnett(2001)]%
        {arnett2001adolescence}
\bibfield{author}{\bibinfo{person}{Jeffrey~Jensen Arnett}.} \bibinfo{year}{2001}\natexlab{}.
\newblock \bibinfo{booktitle}{\emph{Adolescence and emerging adulthood: A cultural approach}}.
\newblock \bibinfo{publisher}{Pearson Education New Zealand}.
\newblock


\bibitem[Arnett and Mitra(2020)]%
        {arnett2020features}
\bibfield{author}{\bibinfo{person}{Jeffrey~Jensen Arnett} {and} \bibinfo{person}{Deeya Mitra}.} \bibinfo{year}{2020}\natexlab{}.
\newblock \showarticletitle{Are the features of emerging adulthood developmentally distinctive? A comparison of ages 18--60 in the United States}.
\newblock \bibinfo{journal}{\emph{Emerging Adulthood}} \bibinfo{volume}{8}, \bibinfo{number}{5} (\bibinfo{year}{2020}), \bibinfo{pages}{412--419}.
\newblock


\bibitem[Baishya and Neustaedter(2017)]%
        {baishya2017inyoureyes}
\bibfield{author}{\bibinfo{person}{Uddipana Baishya} {and} \bibinfo{person}{Carman Neustaedter}.} \bibinfo{year}{2017}\natexlab{}.
\newblock \showarticletitle{In Your Eyes: Anytime, Anywhere Video and Audio Streaming for Couples}. In \bibinfo{booktitle}{\emph{Proceedings of the 2017 ACM Conference on Computer Supported Cooperative Work and Social Computing}} (Portland, Oregon, USA) \emph{(\bibinfo{series}{CSCW '17})}. \bibinfo{publisher}{Association for Computing Machinery}, \bibinfo{address}{New York, NY, USA}, \bibinfo{pages}{84–97}.
\newblock
\showISBNx{9781450343350}
\urldef\tempurl%
\url{https://doi.org/10.1145/2998181.2998200}
\showURL{%
\tempurl}


\bibitem[Bales et~al\mbox{.}(2011)]%
        {bales2011couplevibe}
\bibfield{author}{\bibinfo{person}{Elizabeth Bales}, \bibinfo{person}{Kevin~A Li}, {and} \bibinfo{person}{William Griwsold}.} \bibinfo{year}{2011}\natexlab{}.
\newblock \showarticletitle{CoupleVIBE: mobile implicit communication to improve awareness for (long-distance) couples}. In \bibinfo{booktitle}{\emph{Proceedings of the ACM 2011 conference on Computer supported cooperative work}}. \bibinfo{pages}{65--74}.
\newblock


\bibitem[Ballagas et~al\mbox{.}(2013)]%
        {ballagas2012reading}
\bibfield{author}{\bibinfo{person}{Rafael Ballagas}, \bibinfo{person}{Joseph `Jofish'~Kaye}, {and} \bibinfo{person}{Hayes Raffle}.} \bibinfo{year}{2013}\natexlab{}.
\newblock \bibinfo{booktitle}{\emph{Reading, Laughing, and Connecting with Young Children}}.
\newblock \bibinfo{publisher}{Springer London}, \bibinfo{address}{London}, \bibinfo{pages}{159--172}.
\newblock
\urldef\tempurl%
\url{https://doi.org/10.1007/978-1-4471-4192-1_9}
\showURL{%
\tempurl}


\bibitem[Ballagas et~al\mbox{.}(2009)]%
        {ballagas2009family}
\bibfield{author}{\bibinfo{person}{Rafael Ballagas}, \bibinfo{person}{Joseph'Jofish' Kaye}, \bibinfo{person}{Morgan Ames}, \bibinfo{person}{Janet Go}, {and} \bibinfo{person}{Hayes Raffle}.} \bibinfo{year}{2009}\natexlab{}.
\newblock \showarticletitle{Family communication: phone conversations with children}. In \bibinfo{booktitle}{\emph{Proceedings of the 8th international Conference on Interaction Design and Children}}. \bibinfo{pages}{321--324}.
\newblock


\bibitem[Bassoli et~al\mbox{.}(2006)]%
        {bassoli2006tuna}
\bibfield{author}{\bibinfo{person}{Arianna Bassoli}, \bibinfo{person}{Julian Moore}, {and} \bibinfo{person}{Stefan Agamanolis}.} \bibinfo{year}{2006}\natexlab{}.
\newblock \showarticletitle{tunA: Socialising music sharing on the move}.
\newblock \bibinfo{journal}{\emph{Consuming music together: Social and collaborative aspects of music consumption technologies}} (\bibinfo{year}{2006}), \bibinfo{pages}{151--172}.
\newblock


\bibitem[Bly et~al\mbox{.}(1993)]%
        {bly1993media}
\bibfield{author}{\bibinfo{person}{Sara~A. Bly}, \bibinfo{person}{Steve~R. Harrison}, {and} \bibinfo{person}{Susan Irwin}.} \bibinfo{year}{1993}\natexlab{}.
\newblock \showarticletitle{Media spaces: bringing people together in a video, audio, and computing environment}.
\newblock \bibinfo{journal}{\emph{Commun. ACM}} \bibinfo{volume}{36}, \bibinfo{number}{1} (\bibinfo{date}{Jan.} \bibinfo{year}{1993}), \bibinfo{pages}{28–46}.
\newblock
\showISSN{0001-0782}
\urldef\tempurl%
\url{https://doi.org/10.1145/151233.151235}
\showURL{%
\tempurl}


\bibitem[Boer and Abubakar(2014)]%
        {boer2014music}
\bibfield{author}{\bibinfo{person}{Diana Boer} {and} \bibinfo{person}{Amina Abubakar}.} \bibinfo{year}{2014}\natexlab{}.
\newblock \showarticletitle{Music listening in families and peer groups: benefits for young people's social cohesion and emotional well-being across four cultures}.
\newblock \bibinfo{journal}{\emph{Frontiers in psychology}}  \bibinfo{volume}{5} (\bibinfo{year}{2014}), \bibinfo{pages}{392}.
\newblock


\bibitem[Boer and Fischer(2012)]%
        {boer2012towards}
\bibfield{author}{\bibinfo{person}{Diana Boer} {and} \bibinfo{person}{Ronald Fischer}.} \bibinfo{year}{2012}\natexlab{}.
\newblock \showarticletitle{Towards a holistic model of functions of music listening across cultures: A culturally decentred qualitative approach}.
\newblock \bibinfo{journal}{\emph{Psychology of Music}} \bibinfo{volume}{40}, \bibinfo{number}{2} (\bibinfo{year}{2012}), \bibinfo{pages}{179--200}.
\newblock


\bibitem[Boer et~al\mbox{.}(2011)]%
        {boer2011shared}
\bibfield{author}{\bibinfo{person}{Diana Boer}, \bibinfo{person}{Ronald Fischer}, \bibinfo{person}{Micha Strack}, \bibinfo{person}{Michael~H Bond}, \bibinfo{person}{Eva Lo}, {and} \bibinfo{person}{Jason Lam}.} \bibinfo{year}{2011}\natexlab{}.
\newblock \showarticletitle{How shared preferences in music create bonds between people: Values as the missing link}.
\newblock \bibinfo{journal}{\emph{Personality and Social Psychology Bulletin}} \bibinfo{volume}{37}, \bibinfo{number}{9} (\bibinfo{year}{2011}), \bibinfo{pages}{1159--1171}.
\newblock


\bibitem[Boer et~al\mbox{.}(2012)]%
        {boer2012young}
\bibfield{author}{\bibinfo{person}{Diana Boer}, \bibinfo{person}{Ronald Fischer}, \bibinfo{person}{Hasan~G{\"u}rkan Tekman}, \bibinfo{person}{Amina Abubakar}, \bibinfo{person}{Jane Njenga}, {and} \bibinfo{person}{Markus Zenger}.} \bibinfo{year}{2012}\natexlab{}.
\newblock \showarticletitle{Young people's topography of musical functions: Personal, social and cultural experiences with music across genders and six societies}.
\newblock \bibinfo{journal}{\emph{International Journal of Psychology}} \bibinfo{volume}{47}, \bibinfo{number}{5} (\bibinfo{year}{2012}), \bibinfo{pages}{355--369}.
\newblock


\bibitem[Braun and Clarke(2006)]%
        {braun2006using}
\bibfield{author}{\bibinfo{person}{Virginia Braun} {and} \bibinfo{person}{Victoria Clarke}.} \bibinfo{year}{2006}\natexlab{}.
\newblock \showarticletitle{Using thematic analysis in psychology}.
\newblock \bibinfo{journal}{\emph{Qualitative research in psychology}} \bibinfo{volume}{3}, \bibinfo{number}{2} (\bibinfo{year}{2006}), \bibinfo{pages}{77--101}.
\newblock


\bibitem[Brown et~al\mbox{.}(2001)]%
        {brown2001music}
\bibfield{author}{\bibinfo{person}{Barry Brown}, \bibinfo{person}{Abigail~J Sellen}, {and} \bibinfo{person}{Erik Geelhoed}.} \bibinfo{year}{2001}\natexlab{}.
\newblock \showarticletitle{Music sharing as a computer supported collaborative application}. In \bibinfo{booktitle}{\emph{ECSCW 2001: Proceedings of the Seventh European Conference on Computer Supported Cooperative Work 16--20 September 2001, Bonn, Germany}}. Springer, \bibinfo{pages}{179--198}.
\newblock


\bibitem[Bunker et~al\mbox{.}(1992)]%
        {bunker1992quality}
\bibfield{author}{\bibinfo{person}{Barbara~B Bunker}, \bibinfo{person}{Josephine~M Zubek}, \bibinfo{person}{Virginia~J Vanderslice}, {and} \bibinfo{person}{Robert~W Rice}.} \bibinfo{year}{1992}\natexlab{}.
\newblock \showarticletitle{Quality of life in dual-career families: Commuting versus single-residence couples}.
\newblock \bibinfo{journal}{\emph{Journal of Marriage and the Family}} (\bibinfo{year}{1992}), \bibinfo{pages}{399--407}.
\newblock


\bibitem[Caraban et~al\mbox{.}(2019)]%
        {caraban201923}
\bibfield{author}{\bibinfo{person}{Ana Caraban}, \bibinfo{person}{Evangelos Karapanos}, \bibinfo{person}{Daniel Gon{\c{c}}alves}, {and} \bibinfo{person}{Pedro Campos}.} \bibinfo{year}{2019}\natexlab{}.
\newblock \showarticletitle{23 ways to nudge: A review of technology-mediated nudging in human-computer interaction}. In \bibinfo{booktitle}{\emph{Proceedings of the 2019 CHI conference on human factors in computing systems}}. \bibinfo{pages}{1--15}.
\newblock


\bibitem[Chowdhury(2022)]%
        {chowdhury2022music}
\bibfield{author}{\bibinfo{person}{Nabila Chowdhury}.} \bibinfo{year}{2022}\natexlab{}.
\newblock \emph{\bibinfo{title}{Music co-listening over video chat to support intergenerational connectedness}}.
\newblock \bibinfo{thesistype}{Master's\ thesis}.
\newblock


\bibitem[Chowdhury et~al\mbox{.}(2021)]%
        {chowdhury2021listening}
\bibfield{author}{\bibinfo{person}{Nabila Chowdhury}, \bibinfo{person}{Celine Latulipe}, {and} \bibinfo{person}{James~E Young}.} \bibinfo{year}{2021}\natexlab{}.
\newblock \showarticletitle{Listening Together while Apart: Intergenerational Music Listening}. In \bibinfo{booktitle}{\emph{Companion Publication of the 2021 Conference on Computer Supported Cooperative Work and Social Computing}}. \bibinfo{pages}{36--39}.
\newblock


\bibitem[Elbir et~al\mbox{.}(2018)]%
        {elbir2018music}
\bibfield{author}{\bibinfo{person}{Ahmet Elbir}, \bibinfo{person}{Hilmi~Bilal {\c{C}}am}, \bibinfo{person}{Mehmet~Emre Iyican}, \bibinfo{person}{Berkay {\"O}zt{\"u}rk}, {and} \bibinfo{person}{Nizamettin Aydin}.} \bibinfo{year}{2018}\natexlab{}.
\newblock \showarticletitle{Music genre classification and recommendation by using machine learning techniques}. In \bibinfo{booktitle}{\emph{2018 Innovations in intelligent systems and applications conference (ASYU)}}. IEEE, \bibinfo{pages}{1--5}.
\newblock


\bibitem[Fiese et~al\mbox{.}(2002)]%
        {fiese2002review}
\bibfield{author}{\bibinfo{person}{Barbara~H Fiese}, \bibinfo{person}{Thomas~J Tomcho}, \bibinfo{person}{Michael Douglas}, \bibinfo{person}{Kimberly Josephs}, \bibinfo{person}{Scott Poltrock}, {and} \bibinfo{person}{Tim Baker}.} \bibinfo{year}{2002}\natexlab{}.
\newblock \showarticletitle{A review of 50 years of research on naturally occurring family routines and rituals: Cause for celebration?}
\newblock \bibinfo{journal}{\emph{Journal of family psychology}} \bibinfo{volume}{16}, \bibinfo{number}{4} (\bibinfo{year}{2002}), \bibinfo{pages}{381}.
\newblock


\bibitem[Flanagan et~al\mbox{.}(1993)]%
        {flanagan1993residential}
\bibfield{author}{\bibinfo{person}{Constance Flanagan}, \bibinfo{person}{John Schulenberg}, {and} \bibinfo{person}{Andrew Fuligni}.} \bibinfo{year}{1993}\natexlab{}.
\newblock \showarticletitle{Residential setting and parent-adolescent relationships during the college years}.
\newblock \bibinfo{journal}{\emph{Journal of youth and adolescence}}  \bibinfo{volume}{22} (\bibinfo{year}{1993}), \bibinfo{pages}{171--189}.
\newblock


\bibitem[Fogg(2009)]%
        {fogg2009behavior}
\bibfield{author}{\bibinfo{person}{Brian~J Fogg}.} \bibinfo{year}{2009}\natexlab{}.
\newblock \showarticletitle{A behavior model for persuasive design}. In \bibinfo{booktitle}{\emph{Proceedings of the 4th international Conference on Persuasive Technology}}. \bibinfo{pages}{1--7}.
\newblock


\bibitem[Follmer et~al\mbox{.}(2012)]%
        {follmer2012people}
\bibfield{author}{\bibinfo{person}{Sean Follmer}, \bibinfo{person}{Rafael Ballagas}, \bibinfo{person}{Hayes Raffle}, \bibinfo{person}{Mirjana Spasojevic}, {and} \bibinfo{person}{Hiroshi Ishii}.} \bibinfo{year}{2012}\natexlab{}.
\newblock \showarticletitle{People in books: using a FlashCam to become part of an interactive book for connected reading}. In \bibinfo{booktitle}{\emph{Proceedings of the ACM 2012 conference on Computer supported cooperative work}}. \bibinfo{pages}{685--694}.
\newblock


\bibitem[Gaggioli et~al\mbox{.}(2014)]%
        {gaggioli2014intergenerational}
\bibfield{author}{\bibinfo{person}{Andrea Gaggioli}, \bibinfo{person}{Luca Morganti}, \bibinfo{person}{Silvio Bonfiglio}, \bibinfo{person}{Chiara Scaratti}, \bibinfo{person}{Pietro Cipresso}, \bibinfo{person}{Silvia Serino}, {and} \bibinfo{person}{Giuseppe Riva}.} \bibinfo{year}{2014}\natexlab{}.
\newblock \showarticletitle{Intergenerational group reminiscence: A potentially effective intervention to enhance elderly psychosocial wellbeing and to improve children's perception of aging}.
\newblock \bibinfo{journal}{\emph{Educational Gerontology}} \bibinfo{volume}{40}, \bibinfo{number}{7} (\bibinfo{year}{2014}), \bibinfo{pages}{486--498}.
\newblock


\bibitem[Gamberini et~al\mbox{.}(2007)]%
        {gamberini2007embedded}
\bibfield{author}{\bibinfo{person}{Luciano Gamberini}, \bibinfo{person}{Giovanni Petrucci}, \bibinfo{person}{Andrea Spoto}, {and} \bibinfo{person}{Anna Spagnolli}.} \bibinfo{year}{2007}\natexlab{}.
\newblock \showarticletitle{Embedded persuasive strategies to obtain visitors’ data: Comparing reward and reciprocity in an amateur, knowledge-based website}. In \bibinfo{booktitle}{\emph{Persuasive Technology: Second International Conference on Persuasive Technology, PERSUASIVE 2007, Palo Alto, CA, USA, April 26-27, 2007, Revised Selected Papers 2}}. Springer, \bibinfo{pages}{187--198}.
\newblock


\bibitem[Garcia~Mendoza et~al\mbox{.}(2019)]%
        {garcia2019role}
\bibfield{author}{\bibinfo{person}{Maria Del~Carmen Garcia~Mendoza}, \bibinfo{person}{Inmaculada Sanchez~Queija}, {and} \bibinfo{person}{Agueda Parra~Jimenez}.} \bibinfo{year}{2019}\natexlab{}.
\newblock \showarticletitle{The role of parents in emerging adults’ psychological well-being: A person-oriented approach}.
\newblock \bibinfo{journal}{\emph{Family process}} \bibinfo{volume}{58}, \bibinfo{number}{4} (\bibinfo{year}{2019}), \bibinfo{pages}{954--971}.
\newblock


\bibitem[Golish(2000)]%
        {golish2000changes}
\bibfield{author}{\bibinfo{person}{Tamara~D Golish}.} \bibinfo{year}{2000}\natexlab{}.
\newblock \showarticletitle{Changes in closeness between adult children and their parents: A turning point analysis}.
\newblock \bibinfo{journal}{\emph{Communication Reports}} \bibinfo{volume}{13}, \bibinfo{number}{2} (\bibinfo{year}{2000}), \bibinfo{pages}{79--97}.
\newblock


\bibitem[Gui et~al\mbox{.}(2017)]%
        {gui2017fitness}
\bibfield{author}{\bibinfo{person}{Xinning Gui}, \bibinfo{person}{Yu Chen}, \bibinfo{person}{Clara Caldeira}, \bibinfo{person}{Dan Xiao}, {and} \bibinfo{person}{Yunan Chen}.} \bibinfo{year}{2017}\natexlab{}.
\newblock \showarticletitle{When fitness meets social networks: Investigating fitness tracking and social practices on werun}. In \bibinfo{booktitle}{\emph{Proceedings of the 2017 CHI conference on human factors in computing systems}}. \bibinfo{pages}{1647--1659}.
\newblock


\bibitem[Hakansson et~al\mbox{.}(2007)]%
        {hakansson2007facilitating}
\bibfield{author}{\bibinfo{person}{Maria Hakansson}, \bibinfo{person}{Mattias Rost}, \bibinfo{person}{Mattias Jacobsson}, {and} \bibinfo{person}{Lars~Erik Holmquist}.} \bibinfo{year}{2007}\natexlab{}.
\newblock \showarticletitle{Facilitating mobile music sharing and social interaction with Push! Music}. In \bibinfo{booktitle}{\emph{2007 40th Annual Hawaii International Conference on System Sciences (HICSS'07)}}. IEEE, \bibinfo{pages}{87--87}.
\newblock


\bibitem[Heshmat et~al\mbox{.}(2020)]%
        {heshmat2020familystories}
\bibfield{author}{\bibinfo{person}{Yasamin Heshmat}, \bibinfo{person}{Carman Neustaedter}, \bibinfo{person}{Kyle McCaffrey}, \bibinfo{person}{William Odom}, \bibinfo{person}{Ron Wakkary}, {and} \bibinfo{person}{Zikun Yang}.} \bibinfo{year}{2020}\natexlab{}.
\newblock \showarticletitle{Familystories: Asynchronous audio storytelling for family members across time zones}. In \bibinfo{booktitle}{\emph{Proceedings of the 2020 chi conference on human factors in computing systems}}. \bibinfo{pages}{1--14}.
\newblock


\bibitem[Hindus et~al\mbox{.}(2001)]%
        {hindus2001casablanca}
\bibfield{author}{\bibinfo{person}{Debby Hindus}, \bibinfo{person}{Scott~D. Mainwaring}, \bibinfo{person}{Nicole Leduc}, \bibinfo{person}{Anna~Elizabeth Hagstr\"{o}m}, {and} \bibinfo{person}{Oliver Bayley}.} \bibinfo{year}{2001}\natexlab{}.
\newblock \showarticletitle{Casablanca: designing social communication devices for the home}. In \bibinfo{booktitle}{\emph{Proceedings of the SIGCHI Conference on Human Factors in Computing Systems}} (Seattle, Washington, USA) \emph{(\bibinfo{series}{CHI '01})}. \bibinfo{publisher}{Association for Computing Machinery}, \bibinfo{address}{New York, NY, USA}, \bibinfo{pages}{325–332}.
\newblock
\showISBNx{1581133278}
\urldef\tempurl%
\url{https://doi.org/10.1145/365024.383749}
\showURL{%
\tempurl}


\bibitem[Hofer and Moore(2011)]%
        {hofer2011iconnected}
\bibfield{author}{\bibinfo{person}{Barbara~K Hofer} {and} \bibinfo{person}{Abigail~Sullivan Moore}.} \bibinfo{year}{2011}\natexlab{}.
\newblock \bibinfo{booktitle}{\emph{The iConnected parent: Staying close to your kids in college (and beyond) while letting them grow up}}.
\newblock \bibinfo{publisher}{Simon and Schuster}.
\newblock


\bibitem[Jin et~al\mbox{.}(2023)]%
        {jin2023socio}
\bibfield{author}{\bibinfo{person}{Qiao Jin}, \bibinfo{person}{Ye Yuan}, {and} \bibinfo{person}{Svetlana Yarosh}.} \bibinfo{year}{2023}\natexlab{}.
\newblock \showarticletitle{Socio-technical opportunities in long-distance communication between siblings with a large age difference}. In \bibinfo{booktitle}{\emph{Proceedings of the 2023 CHI Conference on Human Factors in Computing Systems}}. \bibinfo{pages}{1--15}.
\newblock


\bibitem[Judge et~al\mbox{.}(2011)]%
        {judge2011familyportals}
\bibfield{author}{\bibinfo{person}{Tejinder~K. Judge}, \bibinfo{person}{Carman Neustaedter}, \bibinfo{person}{Steve Harrison}, {and} \bibinfo{person}{Andrew Blose}.} \bibinfo{year}{2011}\natexlab{}.
\newblock \showarticletitle{Family portals: connecting families through a multifamily media space}. In \bibinfo{booktitle}{\emph{Proceedings of the SIGCHI Conference on Human Factors in Computing Systems}} (Vancouver, BC, Canada) \emph{(\bibinfo{series}{CHI '11})}. \bibinfo{publisher}{Association for Computing Machinery}, \bibinfo{address}{New York, NY, USA}, \bibinfo{pages}{1205–1214}.
\newblock
\showISBNx{9781450302289}
\urldef\tempurl%
\url{https://doi.org/10.1145/1978942.1979122}
\showURL{%
\tempurl}


\bibitem[Judge et~al\mbox{.}(2010)]%
        {judge2010family}
\bibfield{author}{\bibinfo{person}{Tejinder~K Judge}, \bibinfo{person}{Carman Neustaedter}, {and} \bibinfo{person}{Andrew~F Kurtz}.} \bibinfo{year}{2010}\natexlab{}.
\newblock \showarticletitle{The family window: the design and evaluation of a domestic media space}. In \bibinfo{booktitle}{\emph{Proceedings of the sigchi conference on human factors in computing systems}}. \bibinfo{pages}{2361--2370}.
\newblock


\bibitem[Kang et~al\mbox{.}(2021)]%
        {kang2021momentmeld}
\bibfield{author}{\bibinfo{person}{Bumsoo Kang}, \bibinfo{person}{Seungwoo Kang}, {and} \bibinfo{person}{Inseok Hwang}.} \bibinfo{year}{2021}\natexlab{}.
\newblock \showarticletitle{MomentMeld: AI-augmented Mobile Photographic Memento towards Mutually Stimulatory Inter-generational Interaction}. In \bibinfo{booktitle}{\emph{Proceedings of the 2021 CHI Conference on Human Factors in Computing Systems}}. \bibinfo{pages}{1--16}.
\newblock


\bibitem[Kang et~al\mbox{.}(2017)]%
        {kang2017zaturi}
\bibfield{author}{\bibinfo{person}{Bumsoo Kang}, \bibinfo{person}{Chulhong Min}, \bibinfo{person}{Wonjung Kim}, \bibinfo{person}{Inseok Hwang}, \bibinfo{person}{Chunjong Park}, \bibinfo{person}{Seungchul Lee}, \bibinfo{person}{Sung-Ju Lee}, {and} \bibinfo{person}{Junehwa Song}.} \bibinfo{year}{2017}\natexlab{}.
\newblock \showarticletitle{Zaturi: We put together the 25th hour for you. create a book for your baby}. In \bibinfo{booktitle}{\emph{Proceedings of the 2017 ACM Conference on Computer Supported Cooperative Work and Social Computing}}. \bibinfo{pages}{1850--1863}.
\newblock


\bibitem[Kirk et~al\mbox{.}(2006)]%
        {kirk2006understanding}
\bibfield{author}{\bibinfo{person}{David Kirk}, \bibinfo{person}{Abigail Sellen}, \bibinfo{person}{Carsten Rother}, {and} \bibinfo{person}{Ken Wood}.} \bibinfo{year}{2006}\natexlab{}.
\newblock \showarticletitle{Understanding photowork}. In \bibinfo{booktitle}{\emph{Proceedings of the SIGCHI conference on Human Factors in computing systems}}. \bibinfo{pages}{761--770}.
\newblock


\bibitem[Kirk et~al\mbox{.}(2016)]%
        {kirk2016understanding}
\bibfield{author}{\bibinfo{person}{David~S Kirk}, \bibinfo{person}{Abigail Durrant}, \bibinfo{person}{Gavin Wood}, \bibinfo{person}{Tuck~Wah Leong}, {and} \bibinfo{person}{Peter Wright}.} \bibinfo{year}{2016}\natexlab{}.
\newblock \showarticletitle{Understanding the sociality of experience in mobile music listening with Pocketsong}. In \bibinfo{booktitle}{\emph{Proceedings of the 2016 ACM conference on designing interactive systems}}. \bibinfo{pages}{50--61}.
\newblock


\bibitem[Koerner et~al\mbox{.}(2006)]%
        {koerner2006family}
\bibfield{author}{\bibinfo{person}{Ascan~F Koerner}, \bibinfo{person}{Mary~Anne Fitzpatrick}, {et~al\mbox{.}}} \bibinfo{year}{2006}\natexlab{}.
\newblock \showarticletitle{Family communication patterns theory: A social cognitive approach}.
\newblock \bibinfo{journal}{\emph{Engaging theories in family communication: Multiple perspectives}} (\bibinfo{year}{2006}), \bibinfo{pages}{50--65}.
\newblock


\bibitem[Koropeckyj-Cox(2002)]%
        {koropeckyj2002beyond}
\bibfield{author}{\bibinfo{person}{Tanya Koropeckyj-Cox}.} \bibinfo{year}{2002}\natexlab{}.
\newblock \showarticletitle{Beyond parental status: Psychological well-being in middle and old age}.
\newblock \bibinfo{journal}{\emph{Journal of marriage and family}} \bibinfo{volume}{64}, \bibinfo{number}{4} (\bibinfo{year}{2002}), \bibinfo{pages}{957--971}.
\newblock


\bibitem[Krumhansl and Zupnick(2013)]%
        {krumhansl2013cascading}
\bibfield{author}{\bibinfo{person}{Carol~Lynne Krumhansl} {and} \bibinfo{person}{Justin~Adam Zupnick}.} \bibinfo{year}{2013}\natexlab{}.
\newblock \showarticletitle{Cascading reminiscence bumps in popular music}.
\newblock \bibinfo{journal}{\emph{Psychological science}} \bibinfo{volume}{24}, \bibinfo{number}{10} (\bibinfo{year}{2013}), \bibinfo{pages}{2057--2068}.
\newblock


\bibitem[Kuniavsky(2003)]%
        {kuniavsky2003observing}
\bibfield{author}{\bibinfo{person}{Mike Kuniavsky}.} \bibinfo{year}{2003}\natexlab{}.
\newblock \bibinfo{booktitle}{\emph{Observing the user experience: a practitioner's guide to user research}}.
\newblock \bibinfo{publisher}{Elsevier}.
\newblock


\bibitem[Lee et~al\mbox{.}(2019)]%
        {lee2019can}
\bibfield{author}{\bibinfo{person}{Jin~Ha Lee}, \bibinfo{person}{Liz Pritchard}, {and} \bibinfo{person}{Chris Hubbles}.} \bibinfo{year}{2019}\natexlab{}.
\newblock \showarticletitle{Can We Listen To It Together?: Factors Influencing Reception of Music Recommendations and Post-Recommendation Behavior.}. In \bibinfo{booktitle}{\emph{ISMIR}}. \bibinfo{pages}{663--669}.
\newblock


\bibitem[Leonard(2008)]%
        {leonard2008richard}
\bibfield{author}{\bibinfo{person}{Thomas~C Leonard}.} \bibinfo{year}{2008}\natexlab{}.
\newblock \bibinfo{title}{Richard H. Thaler, Cass R. Sunstein, Nudge: Improving Decisions about Health, Wealth, and Happiness: Yale University Press, New Haven, CT, 2008, 293 pp}.
\newblock
\newblock


\bibitem[Leong and Wright(2013)]%
        {leong2013revisiting}
\bibfield{author}{\bibinfo{person}{Tuck~W Leong} {and} \bibinfo{person}{Peter~C Wright}.} \bibinfo{year}{2013}\natexlab{}.
\newblock \showarticletitle{Revisiting social practices surrounding music}. In \bibinfo{booktitle}{\emph{Proceedings of the SIGCHI conference on human factors in computing systems}}. \bibinfo{pages}{951--960}.
\newblock


\bibitem[Lindell et~al\mbox{.}(2017)]%
        {lindell2017implications}
\bibfield{author}{\bibinfo{person}{Anna~K Lindell}, \bibinfo{person}{Nicole Campione-Barr}, {and} \bibinfo{person}{Sarah~E Killoren}.} \bibinfo{year}{2017}\natexlab{}.
\newblock \showarticletitle{Implications of parent--child relationships for emerging adults’ subjective feelings about adulthood.}
\newblock \bibinfo{journal}{\emph{Journal of Family Psychology}} \bibinfo{volume}{31}, \bibinfo{number}{7} (\bibinfo{year}{2017}), \bibinfo{pages}{810}.
\newblock


\bibitem[Liu and Reimer(2008)]%
        {liu2008social}
\bibfield{author}{\bibinfo{person}{KuanTing Liu} {and} \bibinfo{person}{Roger~Andersson Reimer}.} \bibinfo{year}{2008}\natexlab{}.
\newblock \showarticletitle{Social playlist: enabling touch points and enriching ongoing relationships through collaborative mobile music listening}. In \bibinfo{booktitle}{\emph{Proceedings of the 10th international conference on Human computer interaction with mobile devices and services}}. \bibinfo{pages}{403--406}.
\newblock


\bibitem[Lonsdale and North(2011)]%
        {lonsdale2011we}
\bibfield{author}{\bibinfo{person}{Adam~J Lonsdale} {and} \bibinfo{person}{Adrian~C North}.} \bibinfo{year}{2011}\natexlab{}.
\newblock \showarticletitle{Why do we listen to music? A uses and gratifications analysis}.
\newblock \bibinfo{journal}{\emph{British journal of psychology}} \bibinfo{volume}{102}, \bibinfo{number}{1} (\bibinfo{year}{2011}), \bibinfo{pages}{108--134}.
\newblock


\bibitem[Mantei et~al\mbox{.}(1991)]%
        {mantei1991experiences}
\bibfield{author}{\bibinfo{person}{Marilyn~M. Mantei}, \bibinfo{person}{Ronald~M. Baecker}, \bibinfo{person}{Abigail~J. Sellen}, \bibinfo{person}{William A.~S. Buxton}, \bibinfo{person}{Thomas Milligan}, {and} \bibinfo{person}{Barry Wellman}.} \bibinfo{year}{1991}\natexlab{}.
\newblock \showarticletitle{Experiences in the use of a media space}. In \bibinfo{booktitle}{\emph{Proceedings of the SIGCHI Conference on Human Factors in Computing Systems}} (New Orleans, Louisiana, USA) \emph{(\bibinfo{series}{CHI '91})}. \bibinfo{publisher}{Association for Computing Machinery}, \bibinfo{address}{New York, NY, USA}, \bibinfo{pages}{203–208}.
\newblock
\showISBNx{0897913833}
\urldef\tempurl%
\url{https://doi.org/10.1145/108844.108888}
\showURL{%
\tempurl}


\bibitem[McCurdy et~al\mbox{.}(2022)]%
        {mccurdy2022college}
\bibfield{author}{\bibinfo{person}{Amy~L McCurdy}, \bibinfo{person}{Marta Benito-Gomez}, \bibinfo{person}{Grace~Y Lee}, {and} \bibinfo{person}{Anne~C Fletcher}.} \bibinfo{year}{2022}\natexlab{}.
\newblock \showarticletitle{College Students’ Perceptions of Communication Technology Use and Parent--Child Relationships}.
\newblock \bibinfo{journal}{\emph{Emerging Adulthood}} \bibinfo{volume}{10}, \bibinfo{number}{5} (\bibinfo{year}{2022}), \bibinfo{pages}{1118--1131}.
\newblock


\bibitem[Merz et~al\mbox{.}(2009)]%
        {merz2009wellbeing}
\bibfield{author}{\bibinfo{person}{Eva-Maria Merz}, \bibinfo{person}{Nathan~S Consedine}, \bibinfo{person}{Hans-Joachim Schulze}, {and} \bibinfo{person}{Carlo Schuengel}.} \bibinfo{year}{2009}\natexlab{}.
\newblock \showarticletitle{Wellbeing of adult children and ageing parents: Associations with intergenerational support and relationship quality}.
\newblock \bibinfo{journal}{\emph{Ageing \& Society}} \bibinfo{volume}{29}, \bibinfo{number}{5} (\bibinfo{year}{2009}), \bibinfo{pages}{783--802}.
\newblock


\bibitem[Modlitba(2008)]%
        {modlitba2008globetoddler}
\bibfield{author}{\bibinfo{person}{Lisa~Paulina Modlitba}.} \bibinfo{year}{2008}\natexlab{}.
\newblock \emph{\bibinfo{title}{Globetoddler: Enhancing the experience of remote interaction for preschool children and their traveling parents}}.
\newblock \bibinfo{thesistype}{Ph.\,D. Dissertation}. \bibinfo{school}{Citeseer}.
\newblock


\bibitem[Mynatt et~al\mbox{.}(2001)]%
        {mynatt2001digital}
\bibfield{author}{\bibinfo{person}{Elizabeth~D Mynatt}, \bibinfo{person}{Jim Rowan}, \bibinfo{person}{Sarah Craighill}, {and} \bibinfo{person}{Annie Jacobs}.} \bibinfo{year}{2001}\natexlab{}.
\newblock \showarticletitle{Digital family portraits: supporting peace of mind for extended family members}. In \bibinfo{booktitle}{\emph{Proceedings of the SIGCHI conference on Human factors in computing systems}}. \bibinfo{pages}{333--340}.
\newblock


\bibitem[Neustaedter et~al\mbox{.}(2006)]%
        {neustaedter2006interpersonal}
\bibfield{author}{\bibinfo{person}{Carman Neustaedter}, \bibinfo{person}{Kathryn Elliot}, {and} \bibinfo{person}{Saul Greenberg}.} \bibinfo{year}{2006}\natexlab{}.
\newblock \showarticletitle{Interpersonal awareness in the domestic realm}. In \bibinfo{booktitle}{\emph{Proceedings of the 18th Australia conference on Computer-Human Interaction: Design: Activities, Artefacts and Environments}}. \bibinfo{pages}{15--22}.
\newblock


\bibitem[Nice and Joseph(2023)]%
        {nice2023features}
\bibfield{author}{\bibinfo{person}{Matthew~L Nice} {and} \bibinfo{person}{Matthew Joseph}.} \bibinfo{year}{2023}\natexlab{}.
\newblock \showarticletitle{The features of emerging adulthood and individuation: relations and differences by college-going status, age, and living situation}.
\newblock \bibinfo{journal}{\emph{Emerging Adulthood}} \bibinfo{volume}{11}, \bibinfo{number}{2} (\bibinfo{year}{2023}), \bibinfo{pages}{271--287}.
\newblock


\bibitem[North and Hargreaves(1999)]%
        {north1999music}
\bibfield{author}{\bibinfo{person}{Adrian~C North} {and} \bibinfo{person}{David~J Hargreaves}.} \bibinfo{year}{1999}\natexlab{}.
\newblock \showarticletitle{Music and adolescent identity}.
\newblock \bibinfo{journal}{\emph{Music education research}} \bibinfo{volume}{1}, \bibinfo{number}{1} (\bibinfo{year}{1999}), \bibinfo{pages}{75--92}.
\newblock


\bibitem[O'Hara et~al\mbox{.}(2004)]%
        {ohara2004jukola}
\bibfield{author}{\bibinfo{person}{Kenton O'Hara}, \bibinfo{person}{Matthew Lipson}, \bibinfo{person}{Marcel Jansen}, \bibinfo{person}{Axel Unger}, \bibinfo{person}{Huw Jeffries}, {and} \bibinfo{person}{Peter Macer}.} \bibinfo{year}{2004}\natexlab{}.
\newblock \showarticletitle{Jukola: democratic music choice in a public space}. In \bibinfo{booktitle}{\emph{Proceedings of the 5th Conference on Designing Interactive Systems: Processes, Practices, Methods, and Techniques}} (Cambridge, MA, USA) \emph{(\bibinfo{series}{DIS '04})}. \bibinfo{publisher}{Association for Computing Machinery}, \bibinfo{address}{New York, NY, USA}, \bibinfo{pages}{145–154}.
\newblock
\showISBNx{1581137877}
\urldef\tempurl%
\url{https://doi.org/10.1145/1013115.1013136}
\showURL{%
\tempurl}


\bibitem[Olsson et~al\mbox{.}(2008)]%
        {olsson2008user}
\bibfield{author}{\bibinfo{person}{Thomas Olsson}, \bibinfo{person}{Hannu Soronen}, {and} \bibinfo{person}{Kaisa V{\"a}{\"a}n{\"a}nen-Vainio-Mattila}.} \bibinfo{year}{2008}\natexlab{}.
\newblock \showarticletitle{User needs and design guidelines for mobile services for sharing digital life memories}. In \bibinfo{booktitle}{\emph{Proceedings of the 10th international conference on Human computer interaction with mobile devices and services}}. \bibinfo{pages}{273--282}.
\newblock


\bibitem[Park and Kaneshiro(2021)]%
        {park2021social}
\bibfield{author}{\bibinfo{person}{So~Yeon Park} {and} \bibinfo{person}{Blair Kaneshiro}.} \bibinfo{year}{2021}\natexlab{}.
\newblock \showarticletitle{Social music curation that works: Insights from successful collaborative playlists}.
\newblock \bibinfo{journal}{\emph{Proceedings of the ACM on Human-Computer Interaction}} \bibinfo{volume}{5}, \bibinfo{number}{CSCW1} (\bibinfo{year}{2021}), \bibinfo{pages}{1--27}.
\newblock


\bibitem[Park et~al\mbox{.}(2022)]%
        {park2022cross}
\bibfield{author}{\bibinfo{person}{So~Yeon Park}, \bibinfo{person}{Kyung~Yun Lee}, {and} \bibinfo{person}{Jin~Ha Lee}.} \bibinfo{year}{2022}\natexlab{}.
\newblock \showarticletitle{Cross-Cultural Exploration of Music Sharing}.
\newblock \bibinfo{journal}{\emph{Proceedings of the ACM on Human-Computer Interaction}} \bibinfo{volume}{6}, \bibinfo{number}{CSCW2} (\bibinfo{year}{2022}), \bibinfo{pages}{1--28}.
\newblock


\bibitem[Park et~al\mbox{.}(2012)]%
        {park2012exerlink}
\bibfield{author}{\bibinfo{person}{Taiwoo Park}, \bibinfo{person}{Inseok Hwang}, \bibinfo{person}{Uichin Lee}, \bibinfo{person}{Sunghoon~Ivan Lee}, \bibinfo{person}{Chungkuk Yoo}, \bibinfo{person}{Youngki Lee}, \bibinfo{person}{Hyukjae Jang}, \bibinfo{person}{Sungwon~Peter Choe}, \bibinfo{person}{Souneil Park}, {and} \bibinfo{person}{Junehwa Song}.} \bibinfo{year}{2012}\natexlab{}.
\newblock \showarticletitle{ExerLink: enabling pervasive social exergames with heterogeneous exercise devices}. In \bibinfo{booktitle}{\emph{Proceedings of the 10th international conference on Mobile systems, applications, and services}}. \bibinfo{pages}{15--28}.
\newblock


\bibitem[Priastuty et~al\mbox{.}(2023)]%
        {priastuty2023long}
\bibfield{author}{\bibinfo{person}{Bella Ayu~Dianti Priastuty}, \bibinfo{person}{Salsabila Zahra~Nur Aulia}, \bibinfo{person}{Annida Afifatunnisa}, \bibinfo{person}{Dian Veronika~Sakti Kaloeti}, {et~al\mbox{.}}} \bibinfo{year}{2023}\natexlab{}.
\newblock \showarticletitle{Long-distance, strong connection: Shaping family resilience in the face of long-distance marriage}. In \bibinfo{booktitle}{\emph{Proceedings of International Conference on Psychological Studies (ICPsyche)}}, Vol.~\bibinfo{volume}{4}. \bibinfo{pages}{105--116}.
\newblock


\bibitem[Raffle et~al\mbox{.}(2011)]%
        {raffle2011hello}
\bibfield{author}{\bibinfo{person}{Hayes Raffle}, \bibinfo{person}{Glenda Revelle}, \bibinfo{person}{Koichi Mori}, \bibinfo{person}{Rafael Ballagas}, \bibinfo{person}{Kyle Buza}, \bibinfo{person}{Hiroshi Horii}, \bibinfo{person}{Joseph Kaye}, \bibinfo{person}{Kristin Cook}, \bibinfo{person}{Natalie Freed}, \bibinfo{person}{Janet Go}, {and} \bibinfo{person}{Mirjana Spasojevic}.} \bibinfo{year}{2011}\natexlab{}.
\newblock \showarticletitle{Hello, is grandma there? let's read! StoryVisit: family video chat and connected e-books}. In \bibinfo{booktitle}{\emph{Proceedings of the SIGCHI Conference on Human Factors in Computing Systems}} (Vancouver, BC, Canada) \emph{(\bibinfo{series}{CHI '11})}. \bibinfo{publisher}{Association for Computing Machinery}, \bibinfo{address}{New York, NY, USA}, \bibinfo{pages}{1195–1204}.
\newblock
\showISBNx{9781450302289}
\urldef\tempurl%
\url{https://doi.org/10.1145/1978942.1979121}
\showURL{%
\tempurl}


\bibitem[Rentfrow and Gosling(2006)]%
        {rentfrow2006message}
\bibfield{author}{\bibinfo{person}{Peter~J Rentfrow} {and} \bibinfo{person}{Samuel~D Gosling}.} \bibinfo{year}{2006}\natexlab{}.
\newblock \showarticletitle{Message in a ballad: The role of music preferences in interpersonal perception}.
\newblock \bibinfo{journal}{\emph{Psychological science}} \bibinfo{volume}{17}, \bibinfo{number}{3} (\bibinfo{year}{2006}), \bibinfo{pages}{236--242}.
\newblock


\bibitem[Romero et~al\mbox{.}(2007)]%
        {romero2007connecting}
\bibfield{author}{\bibinfo{person}{Natalia Romero}, \bibinfo{person}{Panos Markopoulos}, \bibinfo{person}{Joy Van~Baren}, \bibinfo{person}{Boris De~Ruyter}, \bibinfo{person}{Wijnand Ijsselsteijn}, {and} \bibinfo{person}{Babak Farshchian}.} \bibinfo{year}{2007}\natexlab{}.
\newblock \showarticletitle{Connecting the family with awareness systems}.
\newblock \bibinfo{journal}{\emph{Personal and Ubiquitous Computing}}  \bibinfo{volume}{11} (\bibinfo{year}{2007}), \bibinfo{pages}{299--312}.
\newblock


\bibitem[Sch{\"a}fer et~al\mbox{.}(2020)]%
        {schafer2020music}
\bibfield{author}{\bibinfo{person}{Katharina Sch{\"a}fer}, \bibinfo{person}{Suvi Saarikallio}, {and} \bibinfo{person}{Tuomas Eerola}.} \bibinfo{year}{2020}\natexlab{}.
\newblock \showarticletitle{Music may reduce loneliness and act as social surrogate for a friend: evidence from an experimental listening study}.
\newblock \bibinfo{journal}{\emph{Music \& Science}}  \bibinfo{volume}{3} (\bibinfo{year}{2020}), \bibinfo{pages}{2059204320935709}.
\newblock


\bibitem[Sch{\"a}fer and Sedlmeier(2009)]%
        {schafer2009functions}
\bibfield{author}{\bibinfo{person}{Thomas Sch{\"a}fer} {and} \bibinfo{person}{Peter Sedlmeier}.} \bibinfo{year}{2009}\natexlab{}.
\newblock \showarticletitle{From the functions of music to music preference}.
\newblock \bibinfo{journal}{\emph{Psychology of Music}} \bibinfo{volume}{37}, \bibinfo{number}{3} (\bibinfo{year}{2009}), \bibinfo{pages}{279--300}.
\newblock


\bibitem[Shakeri et~al\mbox{.}(2023)]%
        {shakeri2023sensing}
\bibfield{author}{\bibinfo{person}{Hanieh Shakeri}, \bibinfo{person}{Denise~Y Geiskkovitch}, \bibinfo{person}{Radhika Garg}, {and} \bibinfo{person}{Carman Neustaedter}.} \bibinfo{year}{2023}\natexlab{}.
\newblock \showarticletitle{Sensing Their Presence: How Emerging Adults And Their Parents Connect After Moving Apart}. In \bibinfo{booktitle}{\emph{Proceedings of the 2023 CHI Conference on Human Factors in Computing Systems}}. \bibinfo{pages}{1--18}.
\newblock


\bibitem[Shin et~al\mbox{.}(2021)]%
        {shin2021designing}
\bibfield{author}{\bibinfo{person}{Ji~Youn Shin}, \bibinfo{person}{Minjin Rheu}, \bibinfo{person}{Jina Huh-Yoo}, {and} \bibinfo{person}{Wei Peng}.} \bibinfo{year}{2021}\natexlab{}.
\newblock \showarticletitle{Designing technologies to support parent-child relationships: a review of current findings and suggestions for future directions}.
\newblock \bibinfo{journal}{\emph{Proceedings of the ACM on Human-Computer Interaction}} \bibinfo{volume}{5}, \bibinfo{number}{CSCW2} (\bibinfo{year}{2021}), \bibinfo{pages}{1--31}.
\newblock


\bibitem[Smith et~al\mbox{.}(2012)]%
        {smith2012going}
\bibfield{author}{\bibinfo{person}{Madeline~E Smith}, \bibinfo{person}{Duyen~T Nguyen}, \bibinfo{person}{Charles Lai}, \bibinfo{person}{Gilly Leshed}, {and} \bibinfo{person}{Eric~PS Baumer}.} \bibinfo{year}{2012}\natexlab{}.
\newblock \showarticletitle{Going to college and staying connected: Communication between college freshmen and their parents}. In \bibinfo{booktitle}{\emph{Proceedings of the ACM 2012 conference on computer supported cooperative work}}. \bibinfo{pages}{789--798}.
\newblock


\bibitem[Stewart et~al\mbox{.}(2018)]%
        {stewart2018co}
\bibfield{author}{\bibinfo{person}{Michael Stewart}, \bibinfo{person}{Javier Tibau}, \bibinfo{person}{Deborah Tatar}, {and} \bibinfo{person}{Steve Harrison}.} \bibinfo{year}{2018}\natexlab{}.
\newblock \showarticletitle{Co-designing for co-listening: Conceptualizing young people’s social and music-listening practices}. In \bibinfo{booktitle}{\emph{Social Computing and Social Media. User Experience and Behavior: 10th International Conference, SCSM 2018, Held as Part of HCI International 2018, Las Vegas, NV, USA, July 15-20, 2018, Proceedings, Part I 10}}. Springer, \bibinfo{pages}{355--374}.
\newblock


\bibitem[Taipale and Farinosi(2018)]%
        {taipale2018big}
\bibfield{author}{\bibinfo{person}{Sakari Taipale} {and} \bibinfo{person}{Manuela Farinosi}.} \bibinfo{year}{2018}\natexlab{}.
\newblock \showarticletitle{The big meaning of small messages: The use of WhatsApp in intergenerational family communication}. In \bibinfo{booktitle}{\emph{Human Aspects of IT for the Aged Population. Acceptance, Communication and Participation: 4th International Conference, ITAP 2018, Held as Part of HCI International 2018, Las Vegas, NV, USA, July 15--20, 2018, Proceedings, Part I 4}}. Springer, \bibinfo{pages}{532--546}.
\newblock


\bibitem[Tang et~al\mbox{.}(2013)]%
        {tang2013homeproxy}
\bibfield{author}{\bibinfo{person}{John~C. Tang}, \bibinfo{person}{Robert Xiao}, \bibinfo{person}{Aaron Hoff}, \bibinfo{person}{Gina Venolia}, \bibinfo{person}{Patrick Therien}, {and} \bibinfo{person}{Asta Roseway}.} \bibinfo{year}{2013}\natexlab{}.
\newblock \showarticletitle{HomeProxy: exploring a physical proxy for video communication in the home}. In \bibinfo{booktitle}{\emph{Proceedings of the SIGCHI Conference on Human Factors in Computing Systems}} (Paris, France) \emph{(\bibinfo{series}{CHI '13})}. \bibinfo{publisher}{Association for Computing Machinery}, \bibinfo{address}{New York, NY, USA}, \bibinfo{pages}{1339–1342}.
\newblock
\showISBNx{9781450318990}
\urldef\tempurl%
\url{https://doi.org/10.1145/2470654.2466175}
\showURL{%
\tempurl}


\bibitem[Tee et~al\mbox{.}(2009)]%
        {tee2009exploring}
\bibfield{author}{\bibinfo{person}{Kimberly Tee}, \bibinfo{person}{AJ~Bernheim Brush}, {and} \bibinfo{person}{Kori~M Inkpen}.} \bibinfo{year}{2009}\natexlab{}.
\newblock \showarticletitle{Exploring communication and sharing between extended families}.
\newblock \bibinfo{journal}{\emph{International Journal of Human-Computer Studies}} \bibinfo{volume}{67}, \bibinfo{number}{2} (\bibinfo{year}{2009}), \bibinfo{pages}{128--138}.
\newblock


\bibitem[Ter~Bogt et~al\mbox{.}(2011)]%
        {ter2011intergenerational}
\bibfield{author}{\bibinfo{person}{Tom~FM Ter~Bogt}, \bibinfo{person}{Marc~JMH Delsing}, \bibinfo{person}{Maarten Van~Zalk}, \bibinfo{person}{Peter~G Christenson}, {and} \bibinfo{person}{Wim~HJ Meeus}.} \bibinfo{year}{2011}\natexlab{}.
\newblock \showarticletitle{Intergenerational continuity of taste: Parental and adolescent music preferences}.
\newblock \bibinfo{journal}{\emph{Social forces}} \bibinfo{volume}{90}, \bibinfo{number}{1} (\bibinfo{year}{2011}), \bibinfo{pages}{297--319}.
\newblock


\bibitem[Tibau et~al\mbox{.}(2019)]%
        {tibau2019familysong}
\bibfield{author}{\bibinfo{person}{Javier Tibau}, \bibinfo{person}{Michael Stewart}, \bibinfo{person}{Steve Harrison}, {and} \bibinfo{person}{Deborah Tatar}.} \bibinfo{year}{2019}\natexlab{}.
\newblock \showarticletitle{FamilySong: Designing to enable music for connection and culture in internationally distributed families}. In \bibinfo{booktitle}{\emph{Proceedings of the 2019 on Designing Interactive Systems Conference}}. \bibinfo{pages}{785--798}.
\newblock


\bibitem[Wallace and Harwood(2018)]%
        {wallace2018associations}
\bibfield{author}{\bibinfo{person}{Sandi~D Wallace} {and} \bibinfo{person}{Jake Harwood}.} \bibinfo{year}{2018}\natexlab{}.
\newblock \showarticletitle{Associations between shared musical engagement and parent--child relational quality: the mediating roles of interpersonal coordination and empathy}.
\newblock \bibinfo{journal}{\emph{Journal of Family Communication}} \bibinfo{volume}{18}, \bibinfo{number}{3} (\bibinfo{year}{2018}), \bibinfo{pages}{202--216}.
\newblock


\bibitem[Wallbaum et~al\mbox{.}(2018)]%
        {wallbaum2018supporting}
\bibfield{author}{\bibinfo{person}{Torben Wallbaum}, \bibinfo{person}{Andrii Matviienko}, \bibinfo{person}{Swamy Ananthanarayan}, \bibinfo{person}{Thomas Olsson}, \bibinfo{person}{Wilko Heuten}, {and} \bibinfo{person}{Susanne~CJ Boll}.} \bibinfo{year}{2018}\natexlab{}.
\newblock \showarticletitle{Supporting communication between grandparents and grandchildren through tangible storytelling systems}. In \bibinfo{booktitle}{\emph{Proceedings of the 2018 CHI Conference on Human Factors in Computing Systems}}. \bibinfo{pages}{1--12}.
\newblock


\bibitem[Xiang et~al\mbox{.}(2020)]%
        {xiang2020links}
\bibfield{author}{\bibinfo{person}{Guangcan Xiang}, \bibinfo{person}{Qingqing Li}, \bibinfo{person}{Xiaoli Du}, \bibinfo{person}{Xinyuan Liu}, \bibinfo{person}{Mingyue Xiao}, {and} \bibinfo{person}{Hong Chen}.} \bibinfo{year}{2020}\natexlab{}.
\newblock \showarticletitle{Links between family cohesion and subjective well-being in adolescents and early adults: The mediating role of self-concept clarity and hope}.
\newblock \bibinfo{journal}{\emph{Current Psychology}} (\bibinfo{year}{2020}), \bibinfo{pages}{1--10}.
\newblock


\bibitem[Yarosh and Abowd(2011)]%
        {yarosh2011mediated}
\bibfield{author}{\bibinfo{person}{Svetlana Yarosh} {and} \bibinfo{person}{Gregory~D Abowd}.} \bibinfo{year}{2011}\natexlab{}.
\newblock \showarticletitle{Mediated parent-child contact in work-separated families}. In \bibinfo{booktitle}{\emph{Proceedings of the SIGCHI conference on human factors in computing systems}}. \bibinfo{pages}{1185--1194}.
\newblock


\bibitem[Yarosh et~al\mbox{.}(2009a)]%
        {yarosh2009supporting}
\bibfield{author}{\bibinfo{person}{Svetlana Yarosh}, \bibinfo{person}{Yee Chieh}, \bibinfo{person}{Gregory~D Abowd}, {et~al\mbox{.}}} \bibinfo{year}{2009}\natexlab{a}.
\newblock \showarticletitle{Supporting parent--child communication in divorced families}.
\newblock \bibinfo{journal}{\emph{International Journal of Human-Computer Studies}} \bibinfo{volume}{67}, \bibinfo{number}{2} (\bibinfo{year}{2009}), \bibinfo{pages}{192--203}.
\newblock


\bibitem[Yarosh et~al\mbox{.}(2009b)]%
        {yarosh2009developing}
\bibfield{author}{\bibinfo{person}{Svetlana Yarosh}, \bibinfo{person}{Stephen Cuzzort}, \bibinfo{person}{Hendrik M\"{u}ller}, {and} \bibinfo{person}{Gregory~D. Abowd}.} \bibinfo{year}{2009}\natexlab{b}.
\newblock \showarticletitle{Developing a media space for remote synchronous parent-child interaction}. In \bibinfo{booktitle}{\emph{Proceedings of the 8th International Conference on Interaction Design and Children}} (Como, Italy) \emph{(\bibinfo{series}{IDC '09})}. \bibinfo{publisher}{Association for Computing Machinery}, \bibinfo{address}{New York, NY, USA}, \bibinfo{pages}{97–105}.
\newblock
\showISBNx{9781605583952}
\urldef\tempurl%
\url{https://doi.org/10.1145/1551788.1551806}
\showURL{%
\tempurl}


\bibitem[Yarosh et~al\mbox{.}(2013)]%
        {yarosh2013almost}
\bibfield{author}{\bibinfo{person}{Svetlana Yarosh}, \bibinfo{person}{Anthony Tang}, \bibinfo{person}{Sanika Mokashi}, {and} \bibinfo{person}{Gregory~D. Abowd}.} \bibinfo{year}{2013}\natexlab{}.
\newblock \showarticletitle{"almost touching": parent-child remote communication using the sharetable system}. In \bibinfo{booktitle}{\emph{Proceedings of the 2013 Conference on Computer Supported Cooperative Work}} (San Antonio, Texas, USA) \emph{(\bibinfo{series}{CSCW '13})}. \bibinfo{publisher}{Association for Computing Machinery}, \bibinfo{address}{New York, NY, USA}, \bibinfo{pages}{181–192}.
\newblock
\showISBNx{9781450313315}
\urldef\tempurl%
\url{https://doi.org/10.1145/2441776.2441798}
\showURL{%
\tempurl}


\bibitem[Yin et~al\mbox{.}(2024)]%
        {yin2024methodological}
\bibfield{author}{\bibinfo{person}{Ines~Ziyou Yin}, \bibinfo{person}{Izzy~Yi Jian}, {and} \bibinfo{person}{Kin Wai~Michael Siu}.} \bibinfo{year}{2024}\natexlab{}.
\newblock \showarticletitle{Methodological considerations of technology co-design with families and design implications on mediating family connectedness from empirical research}.
\newblock \bibinfo{journal}{\emph{Humanities and Social Sciences Communications}} \bibinfo{volume}{11}, \bibinfo{number}{1} (\bibinfo{year}{2024}), \bibinfo{pages}{1--14}.
\newblock


\bibitem[Yuan et~al\mbox{.}(2016)]%
        {yuan2016they}
\bibfield{author}{\bibinfo{person}{Shupei Yuan}, \bibinfo{person}{Syed~A Hussain}, \bibinfo{person}{Kayla~D Hales}, {and} \bibinfo{person}{Shelia~R Cotten}.} \bibinfo{year}{2016}\natexlab{}.
\newblock \showarticletitle{What do they like? Communication preferences and patterns of older adults in the United States: The role of technology}.
\newblock \bibinfo{journal}{\emph{Educational Gerontology}} \bibinfo{volume}{42}, \bibinfo{number}{3} (\bibinfo{year}{2016}), \bibinfo{pages}{163--174}.
\newblock


\end{thebibliography}
